\documentclass{article} 
\usepackage{iclr2023_conference,times}

\usepackage{amsmath,amsfonts,bm}

\def\eqref#1{equation~\ref{#1}}

\def\1{\bm{1}}

\DeclareMathAlphabet{\mathsfit}{\encodingdefault}{\sfdefault}{m}{sl}
\SetMathAlphabet{\mathsfit}{bold}{\encodingdefault}{\sfdefault}{bx}{n}

\usepackage{hyperref}
\usepackage{url}

\iclrfinalcopy 

\usepackage[utf8]{inputenc} 
\usepackage[T1]{fontenc}    
\usepackage{hyperref}       
\usepackage{url}            
\usepackage{amsfonts}       
\usepackage{nicefrac}       
\usepackage{microtype}      
\usepackage{xcolor}         
\usepackage{amsmath}
\usepackage{graphics}
\usepackage{epsfig}
\usepackage{adjustbox}
\usepackage{tabu}
\usepackage{wrapfig,lipsum,booktabs}
\title{Deep Biological Pathway Informed Pathology-Genomic Multimodal Survival Prediction}

\author{%
  Lin Qiu$^1$,  Aminollah Khormali$^2$, Kai Liu$^1$ \\
  $^1$ Division of Research and Early Development, Genentech\\
\{\texttt{qiul13, liuk3}\}\texttt{@gene.com}\\
$^2$  Department of Electrical and Computer Engineering, University of Central Florida \\
   \texttt{aminollah.khormali@gmail.com}
}

\begin{document}

\maketitle

\begin{abstract}
The integration of multi-modal data, such as pathological images and genomic data, is essential for understanding cancer heterogeneity and complexity for personalized treatments, as well as for enhancing survival predictions. Despite the progress made in integrating pathology and genomic data, most existing methods cannot mine the complex inter-modality relations thoroughly. Additionally, identifying explainable features from these models that govern preclinical discovery and clinical prediction is crucial for cancer diagnosis, prognosis, and therapeutic response studies. We propose PONET- a novel biological pathway informed pathology-genomic deep model that integrates pathological images and genomic data not only to improve survival prediction but also to identify genes and pathways that cause different survival rates in patients. Empirical results on six of The Cancer Genome Atlas (TCGA) datasets show that our proposed method achieves superior predictive performance and reveals meaningful biological interpretations. The proposed method establishes insight into how to train biologically informed deep networks on multimodal biomedical data  which will have general applicability for understanding diseases and predicting response and resistance to treatment.

\end{abstract}

\section{Introduction} 
Manual examination of hematoxylin and eosin (H$\&$E)-stained slides of tumor tissue  by pathologists is currently the state-of-the-art for cancer diagnosis \citep{Chan14}. 
The recent advancements in deep learning for digital pathology have enabled the use of whole-slide images (WSIs) for computational image analysis tasks, such as cellular segmentation \citep{Pan17,Hou20}, tissue classification and characterization \citep{hou16,Hekler19,Iizuka20}.
While H$\&$E slides are important and sufficient to establish a profound diagnosis, genomics data can provide a deep molecular characterization of the tumor, potentially offering the chance for prognostic and predictive biomarker discovery.

Cancer prognosis via survival outcome prediction is a standard method used for biomarker discovery, stratification of patients into distinct treatment groups, and therapeutic response prediction \citep{cheng17, ning20}. WSIs exhibit enormous heterogeneity and most approaches adopt a two-stage multiple instance learning-based (MIL) approach for the representation learning of WSIs. Firstly, instance-level feature representations are extracted from image patches in the WSI, and then global aggregation schemes are applied to the bag of instances to obtain a WSI-level representation for subsequent modeling purpose \citep{hou16, courtiol19, wulczyn20, lu21}. Therefore, multimodal survival prediction faces an additional challenge due to the large data heterogeneity gap between WSIs and genomics, and many existing approaches use simple multimodal fusion mechanisms for feature integration, which prevents mining important multimodal interactions \citep{mobadersany18, chen23, chen22}. 

The incorporation of biological pathway databases in a model takes advantage of leveraging prior biological knowledge so that potential prognostic factors of well-known biological functionality can be identified \citep{hao18}. 
Moreover, encoding biological pathway information into the neural networks achieved superior predictive performance compared with established models \citep{elmarakeby21}. 

\begin{figure}[t]
    \centering
    \vskip -0.3in
    \includegraphics[width=0.95\textwidth]{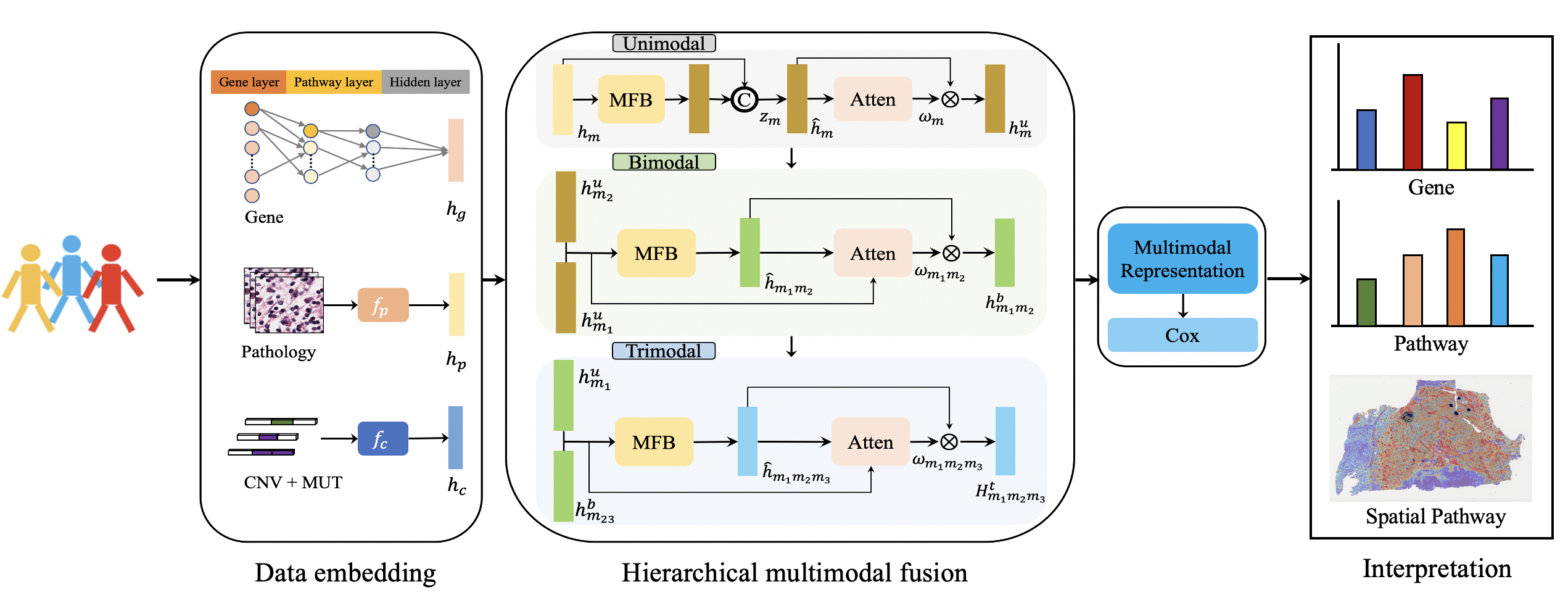}
    \caption{Overview of PONET model.}
    \label{fig:concept}
\end{figure}

Based on the current challenges in multimodal fusion of pathology and genomics and the potential prognostic interpretation to link pathways and clinical outcomes in pathway-based analysis,  we propose a novel biological pathway-informed pathology-genomic deep model, PONET, that uses H$\&$E WSIs and genomic profile features for survival prediction. The proposed method contains four major contributions: 1) PONET formulates a biological pathway-informed deep hierarchical multimodal integration framework for pathological images and genomic data; 2) PONET captures diverse and comprehensive modality-specific and cross-modality relations among different data sources based on the factorized bilinear model and graph fusion network; 3) PONET reveals meaningful model interpretations on both genes and pathways for potential biomarker and therapeutic target discovery; PONET also shows spatial visualization of the top genes/pathways which has enormous potential for novel and prognostic morphological determinants; 4) We evaluate PONET on six public TCGA datasets which showed superior survival prediction comparing to state-of-the-art methods. 
Fig. \ref{fig:concept} shows our model framework.
\section{Related Work}

\textbf{Multimodal Fusion.} Earlier works on multimodal fusion focus on early fusion and late fusion. Early fusion approaches fuse features by simple concatenation which cannot fully explore intra-modality dynamics \citep{wollmer13,poria16,zadeh16}. In contrast, late fusion fuses different modalities by weighted averaging which fails to model cross-modal interactions \citep{Nojavanasghar16,kampman18}. The exploitation of relations within each modality has been successfully introduced in cancer prognosis via bilinear model \citep{wang21} and graph-based model \citep{Subramania21}. Adversarial Representation Graph Fusion (ARGF) \citep{mai20} interprets multimodal fusion as a hierarchical interaction learning procedure where firstly bimodal interactions are generated based on unimodal dynamics, and then trimodal dynamics are generated based on bimodal and unimodal dynamics. We propose a new hierarchical fusion framework with modality-specific and cross-modality attentional factorized bilinear modules to mine the comprehensive modality interactions. Our proposed hierarchical fusion framework is different from ARGF in the following ways: 1) We take the sum of the weighted modality-specific representation as the unimodal representation instead of calculating the weighted average of the modality-specific representation in ARGF; 2) For higher level's fusion, ARGF takes the original embeddings of each modality as input while we use the weighted modality-specific representations; 3) We argue that ARGF takes redundant information during their trimodal dynamics.


\textbf{Multimodal Survival Analysis.} 
There have been exciting attempts on multimodal fusion of pathology and genomic data for cancer survival prediction \citep{mobadersany18, cheerla19, wang19}. However, these multimodal fusion based methods fail to model the interaction between each subset of multiple modalities explicitly. Kronecker product considers pairwise interactions of two input feature vectors by producing a high-dimensional feature of quadratic expansion \citep{zadeh17}, and showed its superiority in cancer survival prediction \citep{wang21, chen23,chen22}. 
Despite promising results, using Kronecker product in multimodal fusion may introduce a large number of parameters that may lead to high computational cost and risk of overfitting \citep{kim17, liu18}, thus limiting its applicability and improvement in performance. To overcome this drawback, hierarchical factorized bilinear fusion for cancer survival prediction (HFBSurv) \citep{li22} uses factorized bilinear model to fuse genomic and image features, dramatically reducing computational complexity.  PONET differs from HFBSurv in two ways: 1) PONET's multimodal framework has three levels of hierarchical fusion module including unimodal, bimodal, and trimodal fusion while HFBSurv only considers within-modality and cross-modality fusion which we argue it is not adequate for mining the comprehensive interactions; 2) PONET leverages biological pathway informed network for better prediction and meaningful interpretation purposes.

\textbf{Pathway-associated Sparse Neural Network.} 
The pathway-based analysis is an approach that a number of studies have investigated to improve both predictive performance and biological interpretability \citep{jin14,cirillo17,hao18, elmarakeby21}. Moreover, pathway-based approaches have shown more reproducible analysis results than gene expression data analysis alone \citep{li15,mallavarapu17}. These pathway-based deep neural networks can only model genomic data which severely inhibits their applicability in current biomedical research. Additionally, the existing pathway-associated sparse neural network structures are limited for disease mechanism investigation: there is only one pathway layer in PASNet \citep{hao18} which contains limited biological prior information to deep dive into the hierarchical pathway and biological process relationships; P-NET \citep{elmarakeby21} calculates the final prediction by taking the average of all the gene and pathway layers' outputs, and this will bias the learning process because it will put more weights for some layers' outputs while underestimating the others. 


\section{Methodology}
\label{gen_inst}

\subsection{Problem formulation and notations}
 The model architecture of PONET is presented in Fig. \ref{fig:concept}, where three modalities are included as input: gene expression $g\in \mathbb{R}^{d_g}$, pathological image $p\in \mathbb{R}^{d_p}$, and copy number (CNV) + mutation (MUT) $CNV+MUT\in \mathbb{R}^{d_c}$, with $d_p$ being the dimensionality of $p$ and so on.  We define a hierarchical factorized bilinear fusion model for PONET.
 We build a sparse biological pathway-informed embedding network for gene expression. A fully connected (FC) embedding layer for both preprocessed pathological image feature ($f_p$) and the copy number + mutation ($f_c$) to map feature into similar embedding space for alleviating the statistical property differences between modalities, the three network architecture details are in Appendix \ref{section:network}. We label the three modality embeddings as $h_m$, $m\in \{g,p,c\}$, the superscript/subscript $u$, $b$, and $t$ represents unimodal fusion, biomodal fusion and trimodal fusion. After that, the embeddings of each modality are first used as input for unimodal fusion to generate the modality-specific representation $h_{m}^u = \omega ^m \hat{h}_{m}$,  $\omega ^m$ represent the modality-specific importance, the feature vector of the unimodal fusion is the sum of all modality-specific representations $h_u = \sum_m h_{m}^u$. In the bimodal fusion, modality-specific representations from the output of unimodal fusion are fused to yield cross-modality representations $h_{m_{1} m_{2}}^{b} = \omega_{m_{1} m_{2}} \hat{h}_{m_{1} m_{2}}, m_{1}, m_{2} \in\{p, c, g\}$ and $m_{1} \neq m_{2}$, $\omega_{m_{1} m_{2}}$ represents the corresponding cross-modality importance. Similarly, the feature vector of bimodal fusion is calculated as $h_b = \sum_{m_{1}, m_{2}} h_{m_{1} m_{2}}^{b}$.  
We propose to build a trimodal fusion to take each cross-modality representation from the output of bimodal fusion to mine the interactions. Similarly to the bimodal fusion architecture, the trimodal fusion feature vector will be $h_t = \sum_{m_{1}, m_{2}, m_{3}} \omega_{m_{1} m_{2} m_{3}} \hat{h}_{m_{1} m_{2} m_{3}},   m_{1}, m_{2}, m_{3} \in \{p, c, g\}$ and $m_{1} \neq m_{2} \neq m_{3}$, $\omega_{m_{1} m_{2} m_{3}}$ represents the corresponding trimodal importance. Finally, PONET concatenates $h_u$, $h_b$, $h_t$ to obtain the final comprehensive multimodal representation and pass it to the Cox proportional hazards model \citep{cox72, cheerla19} for survival prediction. In the following sections we will describe our hierarchical factorized bilinear fusion framework, $l$, $o$, $s$ represents the dimensionality of $h_m$, $z_m$, $\hat{h}_{m1 m2}$.

\subsection{Sparse network}
We design the sparse gene-pathway network consisting of one gene layer followed by three pathway layers. A patient sample of $e$ gene expressions is formed as a column vector, which is denoted by $\mathbf{X} = [x_1, x_2,..., x_e]$, each node represents one gene. 
The gene layer is restricted to have connections reflecting the gene-pathway relationships curated by the Reactome pathway dataset \citep{Jassal20}. The connections are encoded by a binary matrix $\mathbf{M} \in \mathbb{R} ^{a \times e}$, where $a$ is the number of pathways and $e$ is the number of genes, an element of $\mathbf{M}$, $m_{ij}$, is set to one if gene $j$ belongs to pathway $i$. The connections that do not exist in the Reactome pathway dataset will be zero-out. For the following pathway-pathway layers, a similar scheme is applied to control the connection between consecutive layers to reflect the parent-child hierarchical relationships that exist in the Reactome dataset.
The output of each layer is calculated as 
\begin{equation}
  y = f [(\mathbf{M} * \mathbf{W})^T\mathbf{X} + \mathbf{\epsilon}] 
\end{equation}
 where $f$ is the activation function, $\mathbf{M}$ represents the binary matrix, $\mathbf{W}$ is the weights matrix, $\mathbf{X}$ is the input matrix, $\mathbf{\epsilon}$ is the bias vector, and $*$ is the Hadamard product. We use tanh for the activation of each node. We allow the information flow from the biological prior informed network starting from the first gene layer to the last pathway layer, and we label the last layer output embeddings of the sparse network for gene expression as $h_g$.

\subsection{unimodal fusion}

Bilinear models \citep{Tenenbaum00} provide richer representations than linear models. 
Given two feature vectors in different modalities, e.g., the visual features $x\in\mathbb{R}^{m\times 1}$ for an image and the genomic features $y\in\mathbb{R}^{n\times 1}$ for a genomic profile, the bilinear model uses a quadratic expansion of linear transformation considering every pair of features:
\begin{equation}\label{eq:multimodal_bilinear_base}
z_i = x^TW_iy
\end{equation}
where $W_i\in\mathbb{R}^{m\times n}$ is a projection matrix, $z_i\in\mathbb{R}$ is the output of the bilinear model. Bilinear models introduce a large number of parameters which potentially lead to high computational cost and overfitting risk. 
To address these issues, \citet{yu17} develop the Multi-modal Factorized Bilinear pooling (MFB) method, which enjoys the dual benefits of compact output features and robust expressive capacity.

Inspired by the MFB \citep{yu17} and its application in pathology and genomic multimodal learning \citep{li22}, we propose unimodal fusion to capture modality-specific representations and quantify their importance. 
The unimodal fusion takes the embedding of each modality $h_m$ as input and factorizes the projection matrix $W_i$ in Eq. (\ref{eq:multimodal_bilinear_base}) as two low-rank matrices:
\begin{equation}\label{eq:unimodal_fusion_mfb}
\begin{array}{rcl}
z_i &=& h_m^TW_ih_m = \sum\limits_{d=1}^kh_m^Tu_{m,d}v_{m,d}^Th_m\\
&=& 1^T(U_{m,i}^Th_m\circ V_{m,i}^Th_m),  m \in\{p, c, g\}
\end{array}
\end{equation} 
 we get the output feature $z_m$:

\begin{equation}\label{eq:unimodal_z}
z_{m}=\operatorname{SumPooling}\left(\tilde{U}_{m}^{T} h_{m}{ } {\circ} \tilde{V}_{m}^{T} h_{m}, k\right), m \in\{p, c, g\}
\end{equation}

\begin{figure}[t]
    \centering
    \vskip -0.3in
    \includegraphics[width=0.90\textwidth]{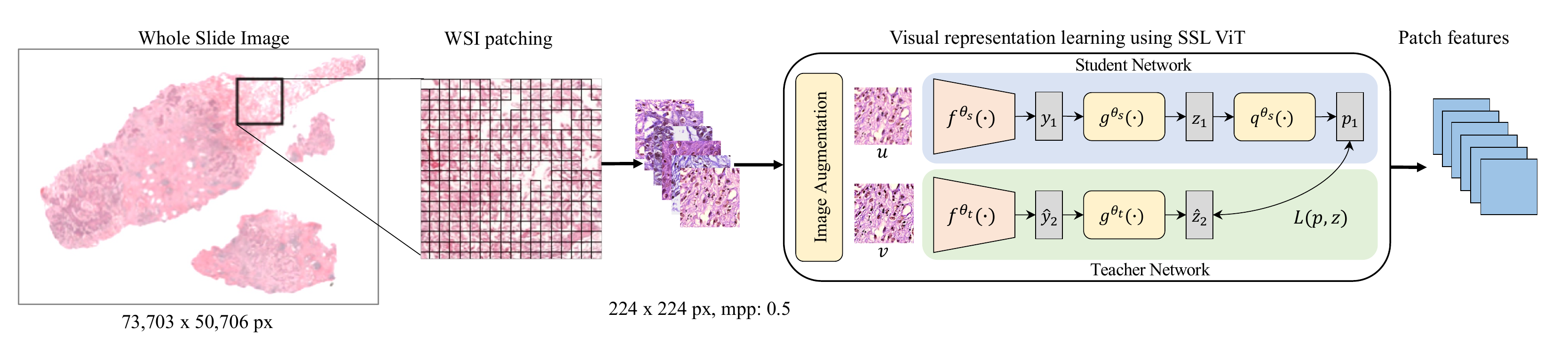}
    \caption{Overall framework of the visual representation extraction using pre-trained self-supervised vision transformer. }
    \label{fig:ViT_Arch}
\end{figure}

where $k$ is the latent dimensionality of the factorized matrices. SumPooling $(x, k)$ function performs sum pooling over $x$ by using a 1-D non-overlapped window with the size $\mathrm{k}, \tilde{U}_{m} \in \mathbb{R}^{l \times k o}$ and $\tilde{V}_{m} \in \mathbb{R}^{l \times k o}$ are 2-D matrices reshaped from $U_{m}$ and $V_{m}$,  $U_{m}=$$\left[U_{m, 1}, \ldots, U_{m, h}\right] \in \mathbb{R}^{l \times k \times o}$ and $V_{m}=\left[V_{m, 1}, \ldots, V_{m, h}\right] \in \mathbb{R}^{l \times k \times o}$. Each modality-specific representation $\hat{h}_{m} \in \mathbb{R}^{l+o}$ is obtained as:
\begin{equation}
\hat{h}_{m}=h_{m} \text {\copyright} z_{m}, m \in\{p, c, g\}
\end{equation}
where $\copyright$ denotes vector concatenation. We also introduce a modality attention network $Atten \in \mathbb{R} ^{l + o} \rightarrow \mathbb{R} ^1 $  to determine the weight for each modality-specific representation to quantify its importance:
\begin{equation}
\omega_{m}=  Atten (\hat{h}_{m};\Theta_{Atten}),  m \in\{p, c, g\}
\end{equation}

where $\omega_m$ is the weight of modality $m$. In practice, $Atten$ consists of a sigmoid activated dense layer parameterized by $\Theta_{Atten}$. 
Therefore, the output of each modality in unimodal fusion, ${h}_{m}^{u}$, is denoted as $\omega_{m} \hat{h}_{m} \in \mathbb{R}^{l+o}, m \in\{p, c, g\}$.  Accordingly, the output of unimodal fusion, $h_u$, is the sum of each weighted modality-specific representation $\omega_{m} \hat{h}_{m}, m \in \{p, c, g\}$ which is different from ARGF \citep{mai20} that used the weighted average of different modalities as the unimodal fusion output.

\subsection{bimodal and trimodal fusion}
Bimodal fusion aims to fuse diverse information of different modalities and quantify different importance for them. After receiving the modality-specific representations ${h}_{m}^{u}$ from the unimodal fusion, we can generate the cross-modality representation $\hat{h}_{m_{1} m_{2}} \in \mathbb{R}^{s}$  similar to Eq. (\ref{eq:unimodal_z}) :
\begin{equation} \label{eq:bimodal_h}
\begin{gathered}
\hat{h}_{m 1, m 2}=\operatorname{Sum} \operatorname{Pooling}\left(\tilde{U}_{m_{1}}^{T} h_{m_1}^{u}{ } {\circ} \tilde{V}_{m_{2}}^{T} h_{m_2}^{u}, k\right), \\
m_{1}, m_{2} \in\{p, c, g\}, m_{1} \neq m_{2}
\end{gathered}
\end{equation}

where $\quad \tilde{U}_{m_1}^{T} \in \mathbb{R}^{(l+o) \times k s}$ and $\tilde{V}_{m_2}^{T} \in \mathbb{R}^{(l+o) \times k s}$ are 2-D matrices reshaped from $U_{m_1}$ and $V_{m_2}$ and $U_{m_1}=\left[U_{m_1,1}, \ldots, U_{m_1, s}\right] \in \mathbb{R}^{(l+o) \times k \times s}$ and $V_{m_2}=\left[V_{m_2,1}, \ldots, V_{m_2, s}\right] \in \mathbb{R}^{(l+o) \times k \times s}$. 
 We leverage a bimodal attention network \citep{mai20} to identify the importance of the cross-modality representation. The similarity $S_{m_{1} m_{2}} \in \mathbb{R}^{1}$ of $ h_{m_{1}}^{u}$ and $h_{m_{2}}^{u}$ is first estimated as follows:
\begin{equation}
S_{m_{1}, m_{2}}=\sum_{i=1}^{l+o}\left(\frac{e^{\omega_{m_{1}}} h_{m_{1}, i}^{u}}{\sum_{j=1}^{l+o} e^{\omega_{m_{1}} h_{m_{1}, j}^{u}}}\right)\left(\frac{e^{\omega_{m_{2}} h_{m_{2}, i}^{u}}}{\sum_{j=1}^{l+o} e^{\omega_{m_{2}} h_{m_{2}, j}^{u}}}\right)
\end{equation}

where the computed similarity is in the range of 0 to 1. 
Then, the cross-modality importance $\omega_{m_{1} m_{2}}$ is obtained by:

\begin{equation}
\omega_{m_{1} m_{2}}=\frac{e^{\hat{\omega}_{m_{i} m_{j}}}}{\sum_{m_{i} \neq m_{j}} e^{\hat{\omega}_{m_{i} m_{j}}}}, \hat{\omega}_{m_{1} m_{2}}=\frac{\omega_{m_{1}}+\omega_{m_{2}}}{S_{m_{1} m_{2}}+S_{0}}
\end{equation}
where $S_{0}$ represents a pre-defined term controlling the relative contribution of similarity and modality-specific importance, and here is set to $0.5$. Therefore, the output of bimodal fusion, $h_b$, is the sum of each weighted cross-modality representation $\omega_{m_{1} m_{2}} \hat{h}_{m_{1} m_{2}}, m_{1}, m_{2} \in \{p, c, g\}$ and $m_{1} \neq m_{2}$. 

In trimodal fusion, each bimodal fusion output is fused with the unimodal fusion output that does not contribute to the formation of the bimodal fusion. The output for each corresponding trimodal representation is $\hat{h}_{m_1 m_2 m_3}$. In addition, trimodal attention was applied to identify the importance of each trimodal representation, $\omega_{m_1 m _2 m_3}$. The output of the trimodal fusion, $h_t$, is the sum of each weighted trimodal representation $\omega_{m_{1} m_{2} m_{3}} \hat{h}_{m_{1} m_{2} m_{3}}, m_{1}, m_{2}, m_{3} \in \{p, c, g\}$ and $m_{1} \neq m_{2} \neq m_{3}$.

\subsection{Survival Loss Function}
We train the model through the Cox partial likelihood loss \citep{cheerla19} with $l_1$ regularization for survival prediction, which is defined as:
\begin{equation}
\ell(\Theta)=-\sum_{i: E_i=1}\left(\hat{\mathfrak{h}}_{\Theta}\left(x_i\right)-\log \sum_{j: T_i>T_j} \exp \left(\hat{\mathfrak{h}}_{\Theta}\left(x_j\right)\right)\right)+\lambda\left(\|\Theta\|_1\right)
\end{equation}
where the values $E_i, T_i$ and $x_i$ for each patient represent the survival status, the survival time, and the feature, respectively. $E_i$ = 1 means event while $E_i$ = 0 represents censor.  $\hat{\mathfrak{h}}_{\Theta}$ is the neural network model trained for predicting the risk of survival, $\Theta$ is the neural network model parameters, and $\lambda$ is a regularization hyperparameter to avoid overfitting.
\section{Experiments}
\label{headings}

\subsection{Experimental Setup}

\textbf{Datasets.} To validate our proposed method, we used six cancer datasets from The Cancer Genome Atlas (TCGA), a public cancer data consortium that contains matched diagnostic WSIs and genomic data with labeled survival times and censorship statuses. The genomic profile features (mutation status, copy number variation,
RNA-Seq expression) are preprocessed by Porpoise \footnote{https://github.com/mahmoodlab/PORPOISE} \citep{chen23}. 
For this study, we used the following cancer types: Bladder Urothelial Carcinoma (BLCA)
(n = 437), Kidney Renal Clear Cell Carcinoma (KIRC) (n =
350), Kidney Renal Papillary Cell Carcinoma (KIRP) (n = 284), Lung Adenocarcinoma (LUAD) (n = 515), Lung Squamous Cell Carcinoma (LUSC) (n = 484), Pancreatic adenocarcinoma (PAAD) (n = 180).  We downloaded the same diagnostic WSIs from the TCGA website \footnote{https://www.cancer.gov/about-nci/organization/ccg/research/structural-genomics/tcga} that were used in Porpoise study to match the paired genomic features and survival times. The feature alignment table for all the cancer types is in Appendix \ref{appendix:data}.
For each WSI, automated segmentation of tissue was performed. Following segmentation, image patches of size 224 $\times$ 224 were extracted without overlap at the 20 X equivalent pyramid level from all tissue regions identified while excluding the white background and selecting only patches with at least 50\% tissue regions. Subsequently, 
a visual representation of those patches is extracted with a vision transformer \citep{wang2021transpath} pre-trained on the TCGA dataset through a self-supervised constructive learning approach, such that each patch is represented as a 1 $\times$ 2048 vector. Fig. \ref{fig:ViT_Arch} shows the framework for the visual representation extraction by vision transformer (VIT). Survival outcome information is available at the patient level, we aggregated the patch-level feature into slide level feature representations based on an attention-based method \citep{lu21, llse18}. 

\textbf{Baselines.} Using the same 5-fold cross-validation splits for evaluating
PONET, we implemented and evaluated six state-of-the-art methods for survival outcome prediction. Additionally, we included three variations of PONET: a) PONET-O represents only genomic data, and pathway architecture for the gene expression are included in the model; b) PONET-OH represents only genomic and pathological image data but without pathway architecture in the model; c) PONET is our full model. For all methods, we use the same VIT feature extraction pipeline for WSIs, as well as identical training hyperparameters and loss functions for supervision. Training details and the parameters tuning can be found in Appendix \ref{appendix:training_details}.

\textbf{CoxPH} \citep{cox72} represents the standard Cox proportional hazard models. \\ 
\textbf{DeepSurv} \citep{katzman18} is the deep neural network version of the CoxPH model. \\
\textbf{Pathomic Fusion} \citep{chen22} as a pioneered deep learning-based framework for predicting survival outcomes by fusing pathology and genomic multimodal data, in which Kronecker product is taken to model pairwise feature interactions across modalities.\\
\textbf{GPDBN} \citep{wang21} adopts Kronecker product to model inter-modality and intra-modality relations between pathology and genomic data for cancer prognosis prediction.\\
\textbf{HFBSurv} \citep{li22} extended GPDBN using the factorized bilinear model to fuse genomic and pathology features in a within-modality and cross-modalities hierarchical fusion. \\
\textbf{Porpoise} \citep{chen23} applied the discrete survival model and Kronecker product to fuse pathology and genomic data for survival prediction \citep{zadeh20}.

 \textbf{Evaluation.}  For each cancer dataset, we  used the cross-validated concordance index (C-Index) (Appendix \ref{appendix:cindx}) \citep{harrell82} to measure the predictive performance of correctly ranking the predicted patient risk scores with respect to overall survival.

\subsection{Results}
\label{results}

\textbf{Comparison with Baselines.} In combing pathology image, genomics, and pathway network via PONET, our approach outperforms CoxPH models, unimodal networks, and previous deep learning-based approaches on pathology-genomic-based survival outcome prediction (Table \ref{table:cindex}). The results show that deep learning-based approaches generally perform better than the CoxPH model. PONET achieves superior C-index values in all six cancer types. All versions of PONET outperform Pathomic Fusion by a big margin. Pathomic Fusion uses Kronecker product to fuse the two modalities, and that's also the reason why other advanced fusion methods, like GPDBN and HFBSurv, achieve better performance. Also, we argue that Pathomic Fusion extracts the region of interest of pathology image for feature extraction might limit the understanding of the tumor microenvironment of the whole slide. HFBSurv shows better performance than GPDBN and Pathomic Fusion which is consistent with their findings, and these results further demonstrate that the hierarchical factorized bilinear model can better mine the rich complementary information among different modalities compared to the Kronecker product. Porpoise performs similarly with PONET on TCGA-KIRC and TCGA-KIRP and outperformed HFBSurv in these two studies, this probably is due to Porpoise partitioned the survival time into different non-overlapping bins and parameterized it as a discrete survival model \citep{zadeh20} which works better for these two cancer types.  In other cases, Porpoise performs similarly to HFBSurv. Note: the results of Porpoise are from their paper \citep{chen23}.

Additionally, we can see that PONET consistently outperforms PONET-O and PONET-OH indicating the effectiveness of the biological pathway-informed neural network and the contribution of pathological image for the overall survival prediction.   
\begin{table}
\centering
\vskip -0.5in
\tiny
\caption{C-Index (mean $\pm$ standard deviation) of PONET and ablation experiments in TCGA survival prediction. The top two performers are highlighted in bold.}
\vspace{0.5em}
\begin{tabular}{l | l | l | l | l| l | l}
\toprule
Model         &   \texttt{TCGA-BLCA} &  \texttt{TCGA-KIRC} &   \texttt{TCGA-KIRP}  &   \texttt{TCGA-LUAD}  &  \texttt{TCGA-LUSC}  & \texttt{TCGA-PAAD}\\
\midrule
CoxPH (Age + Gender) \citep{cox72}  & 0.525 $\pm$ 0.130 & 0.550 $\pm$ 0.070  & 0.544 $\pm$ 0.050 & 0.531 $\pm$ 0.082 & 0.532 $\pm$ 0.094 & 0.539 $\pm$ 0.092\\
\midrule
DeepSurv \citep{kampman18} &   0.580 $\pm$ 0.062 & 0.620  $\pm$ 0.043 & 0.560$\pm$ 0.063 & 0.534 $\pm$0.077 & 0.541 $\pm$ 0.066  &0.544 $\pm$ 0.076\\
\midrule
GPDBN \citep{wang21} & 0.612  $\pm$ 0.042 & 0.647 $\pm$ 0.073 & 0.669 $\pm$ 0.109 & 0.565  $\pm$ 0.057 & 0.545 $\pm$ 0.063 & 0.571 $\pm$ 0.060\\
\midrule
HFBSurv  \citep{li22} & 0.622  $\pm$ 0.043 & 0.667 $\pm$ 0.053 & 0.769 $\pm$ 0.109 & 0.581  $\pm$ 0.017 & 0.548 $\pm$ 0.049 & 0.591 $\pm$ 0.052\\
\midrule
Pathomic Fusion \citep{chen22} & 0.586  $\pm$ 0.062 & 0.598 $\pm$ 0.060 & 0.577 $\pm$ 0.032 & 0.543  $\pm$ 0.065 & 0.523 $\pm$0.045 & 0.545 $\pm$ 0.064  \\

\midrule
Porpoise \citep{chen23} &  0.617  $\pm$ 0.048 & \textbf {0.711 $\pm$ 0.051} & \textbf {0.811 $\pm$ 0.089} & 0.586 $\pm$0.056 & 0.527 $\pm$ 0.043  & 0.591 $\pm$ 0.064 \\

\midrule
\textbf {PONET-O (ours)} & 0.596  $\pm$ 0.056  & 0.664 $\pm$ 0.056 & 0.761 $\pm$ 0.093 & \textbf {0.623 $\pm$0.062} & 0.538 $\pm$ 0.037 & \textbf{0.598 $\pm$ 0.027}\\
\textbf{PONET-OH (ours)}   &  \textbf{0.625 $\pm$ 0.063} & 0.695 $\pm$ 0.043 & 0.776 $\pm$ 0.123 &0.618 $\pm$ 0.049 & \textbf {0.553 $\pm$ 0.049} & 0.591 $\pm$ 0.050 \\
\textbf{PONET (ours)}  &  \textbf {0.643  $\pm$ 0.037} & \textbf {0.726 $\pm$ 0.056} & \textbf {0.829 $\pm$ 0.054} & \textbf {0.646 $\pm$0.047} & \textbf {0.567 $\pm$ 0.066} & \textbf{0.639 $\pm$ 0.080} \\
\bottomrule
\end{tabular}\label{table:cindex}
\end{table}

\textbf{Ablation Studies.} To assess whether the impact of hierarchical factorized bilinear fusion strategy is indeed effective, we compare PONET with four single-fusion methods: 1) Simple concatenation: concatenate each modality embeddings; 2) Element-wise addition: element-wise addition from each modality embeddings; 3) Tensor fusion \citep{zadeh17}: Kronecker product from each modality embeddings. Table \ref{tab:ablation} shows the C-index values of different methods. We can see that PONET achieves the best performance and shows remarkable improvement over single-fusion methods on different cancer type datasets. For example, PONET outperforms the Simple concatenation by 8.4\% (TCGA-BLCA), 27\% (TCGA-KIRP), 15\% (TCGA-LUAD), 8.0\% (TCGA-LUSC), and 11.4\% (TCGA-PAAD), etc. 
\begin{table}[t!]
\centering
\tiny
\caption{Evaluation of PONET on different fusion methods and pathway designs by C-index (mean $\pm$ standard deviation). The best performer is highlighted in bold.}
\begin{tabu}{@{}ccc c c cc@{}}
\toprule
\multicolumn{1}{l}{}                                                                   & Methods    &  \texttt{TCGA-BLCA}         & \texttt{TCGA-KIRP}    & \texttt{TCGA-LUAD}   & \texttt{TCGA-LUSC} & \texttt{TCGA-PAAD}    \\ \midrule
{\begin{tabular}[c]{@{}c@{}}\textsf{Single fusion}\\   \end{tabular}}    & \textsf{Simple concatenation}        & 0.585 $\pm$ 0.045 & 0.652 $\pm$ 0.049 & 0.554 $\pm$ 0.065 & 0.525 $\pm$ 0.066 &  0.568 $\pm$ 0.075  \\ 
                                                                                       & \textsf{Element-wise addition} & 0.592 $\pm$ 0.047    & 0.655 $\pm$ 0.055  &  0.587 $\pm$ 0.065      &   0.522 $\pm$ 0.046   &  0.588 $\pm$ 0.055  \\
                                                                                       & 
                                               \textsf{Tensor fusion \citep{zadeh17}}   & 0.605 $\pm$ 0.046 & 0.775 $\pm$ 0.053 &  0.595 $\pm$ 0.060  &  0.545 $\pm$ 0.045  &  0.592 $\pm$ 0.061   \\                                                                            \midrule

{\begin{tabular}[c]{@{}c@{}}\textsf{Hierarchical fusion}\\       \end{tabular}}      & \textsf{Unimodal}        &  0.596 $\pm$ 0.035  & 0.783 $\pm$ 0.063 & 0.611 $\pm$ 0.056     &  {0.553 $\pm$ 0.073}  & {0.595 $\pm$ 0.053}   \\
                                                                                       & \textsf{Bimodal}  & 0.602 $\pm$ 0.062  & 0.789 $\pm$ 0.053 & 0.601 $\pm$ 0.056     &  {0.552 $\pm$ 0.051}  & {0.598 $\pm$ 0.083}\\
                                                                                       & 
                                                \textsf{ARGF \citep{mai20}}  & 0.597 $\pm$ 0.054 &  0.792 $\pm$ 0.043 &  0.614 $\pm$ 0.051 & 0.556 $\pm$ 0.063 & 0.602 $\pm$ 0.065\\ 
                                                                 &
                                                \textsf{Unimodal + Bimodal}& 0.614 $\pm$ 0.052 & 0.803 $\pm$ 0.061 &  0.631 $\pm$ 0.044  & \textbf{0.578 $\pm$ 0.058} & {0.615 $\pm$ 0.057} \\ 
                                                 \midrule 
                                                 
{\begin{tabular}[c]{@{}c@{}}\textsf{Pathway design}\\       \end{tabular}}      & \textsf{PASNet \citep{hao18}}        &  0.606 $\pm$ 0.045  & 0.793 $\pm$ 0.051 & 0.621 $\pm$ 0.061     &  {0.551 $\pm$ 0.069}  & {0.625 $\pm$ 0.057}   \\
                                                                                       & \textsf{P-NET \citep{elmarakeby21}}  & 0.622 $\pm$ 0.047  & 0.802 $\pm$ 0.071 & 0.625 $\pm$ 0.045     &  {0.562 $\pm$ 0.054}  & {0.627 $\pm$ 0.073}\\
                                                                                       & 
                
                                            \textsf{PONET}&  \textbf{0.643 $\pm$ 0.037} & \textbf{0.829 $\pm$ 0.054} & \textbf{0.641 $\pm$ 0.046} & {0.567 $\pm$ 0.066} & \textbf{0.639 $\pm$ 0.070}  \\                                       
                                                 \bottomrule

\end{tabu}
\vspace{-2em}
\label{tab:ablation}
\end{table}

Furthermore, we adopted five different configurations of PONET to evaluate each hierarchical component of the proposed method: 1) Unimodal: unimodal fusion output as the final feature representation; 2) Bimodal: bimodal fusion output as the final feature representation; 3) Unimodal + Bimodal:
hierarchical (include both unimodal and bimodal feature representation) fusion; 4) ARGF: ARGF \citep{mai20} fusion strategy; 5) PONET: our
proposed hierarchical strategy by incorporating unimodal, bimodal, and trimodal fusion output. As shown in Table \ref{tab:ablation}, Unimodal + Bimodal performs better than Unimodal and Bimodal which demonstrates that Unimodal  + Bimodal can capture the relations within each modality and across modalities. ARGF performs worse than Unimodal + Bimodal and far worse than PONET across all the cancer types. PONET outperforms Unimodal + Bimodal in 4 out of 5 cancer types indicating that three layers of hierarchical fusion can mine the comprehensive interactions among different modalities. 

To evaluate our sparse gene-pathway network design, we compare PONET with PASNet \citep{hao18} and P-NET \citep{elmarakeby21} pathway architecture, PASNet performs the worst due to the fact that it only has one pathway layer in the network, and thus limited prior information was used to predict the outcome. PONET constantly outperforms P-NET across all the cancer types, which demonstrates that averaging all the intermediate layers' output for the final prediction cannot fully capture the prior information flow among the hierarchical biological structures.

\begin{figure}[t!]
    \centering
    \vskip -0.3in
    \includegraphics[width=0.90\textwidth]{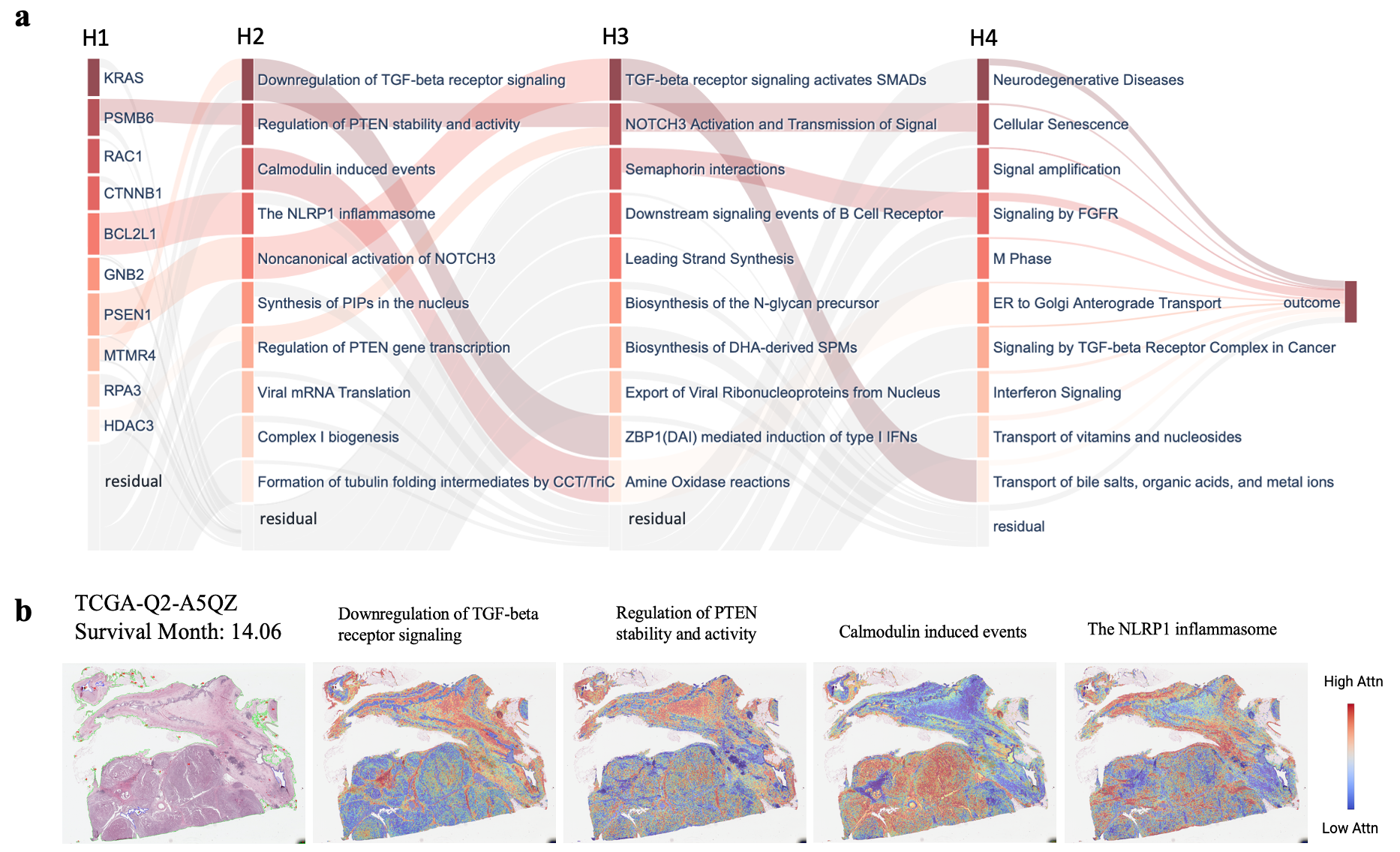}
    \caption{Inspecting and interpreting PONET on TCGA-KIRP. \textbf{a}: Sankey diagram visualization of inner layers of PONET shows the estimated relative importance of different nodes in each layer. Nodes in the first layer represent genes; the next layers represent pathways; and the final layer represents the model outcome. Different layers are linked by weights. Nodes with darker colors are more important, while transparent nodes represent the residual importance of undisplayed nodes in each layer, H1 presents the gene layer, and H2-H4 represent pathway layers; \textbf{b}: Co-attention visualization of top 4 ranked pathways in one case of TCGA-KIRP.}
    \label{fig:sankey}
     \vskip -0.2in
\end{figure}

\textbf{Model Interpretation.} We discuss the model interpretation results for cancer type TCGA-KIRP here and the results for other cancer types are included in the Appendix \ref{appendix:additional_results}. To understand the interactions between different genes, pathways, and biological processes that contributed to the predictive performance and to study the paths of impact from the input to the outcome, we visualized the whole structure of PONET with the fully interpretable layers after training (Fig. \ref{fig:sankey} a). To evaluate the relative importance of specific genes contributing to the model prediction, we inspected the genes layer and used the Integrated Gradients attribution \citep{Sundararajan17} method to obtain the total importance score of genes, and the modified ranking algorithm details are included in the Appendix \ref{appendix:Intgradient}. Highly ranked genes included KRAS, PSMB6, RAC1, and CTNNB1 which are known kidney cancer drivers previously \citep{yang17, shan17,ai-obaidy20, guo22}. GBN2, a member of the guanine nucleotide-binding proteins family, has been reported that the decrease of its expression reduced tumor cell proliferation \citep{zhang19}. A recent study identified a strong dependency  on BCL2L1, which encodes the BCL-XL anti-apoptotic protein, in a subset of kidney cancer cells \citep{grubb22}. This biological interpretability revealed established and novel molecular features contributing to kidney cancer. In addition, PONET selected a hierarchy of pathways relevant to the model prediction, including downregulation of TGF-$\beta$ receptor signaling, regulation of PTEN stability and activity, the NLRP1 inflammasome, and noncanonical activation of NOTCH3 by PSEN1, PSMB6, and BCL2L1. TGF-$\beta$ signaling is increasingly recognized as a key driver in cancer, and in progressive cancer tissues TGF-$\beta$ promotes tumor formation, and its increased expression often correlates with cancer malignancy \citep{han18}. Noncanonical activation of NOTCH3 was reported to limit tumor angiogenesis and plays a vital role in kidney disease \citep{lin17}.

To further inspect the pathway spatial association with the WSI slide we adopted the co-attention survival method MCAT \citep{chen21} between WSIs and genomic features on the top pathways of the second layer, visualized as a WSI-level attention heatmap for each pathway genomic embedding in Fig. \ref{fig:sankey} b (algorithm details are included in the Appendix \ref{appendix:pathway_visu}). We used the gene list from the top 4 pathways as the genomic features and trained MCAT on the TCGA-KIRP dataset for survival prediction. Overall, we observe that high attention in different pathways showed different spatial pattern associations with the slide. This heatmap can reflect genotype-phenotype relationships in cancer pathology. 
The high attention regions (red) of different pathways in the heatmap have positive associations with the predicted death risk while the low attention regions (blue) have negative associations with the predicted risk. By further checking the cell types in high attention patches we can gain insights of prognostic morphological determinants and have a better understanding of the complex tumor microenvironment.
\begin{wraptable}{r}{8cm}
\footnotesize
\caption{Comparison of model complexity}\label{wrap-tab:1}
\begin{tabular}{ccc}\\\toprule  
Methods & Number of Parameters & FLOPS \\\midrule
Pathomic Fusion &175M & 168G\\  \midrule
GPDBN & 82M & 91G\\  \midrule
HFBSurv &0.3M & 0.5G\\  \midrule
PONET & 2.8M & 3.1G\\\bottomrule
\end{tabular}
\end{wraptable}

\begin{figure}[t]
    \centering
    \vskip -0.3in
    \includegraphics[width=0.85\textwidth]{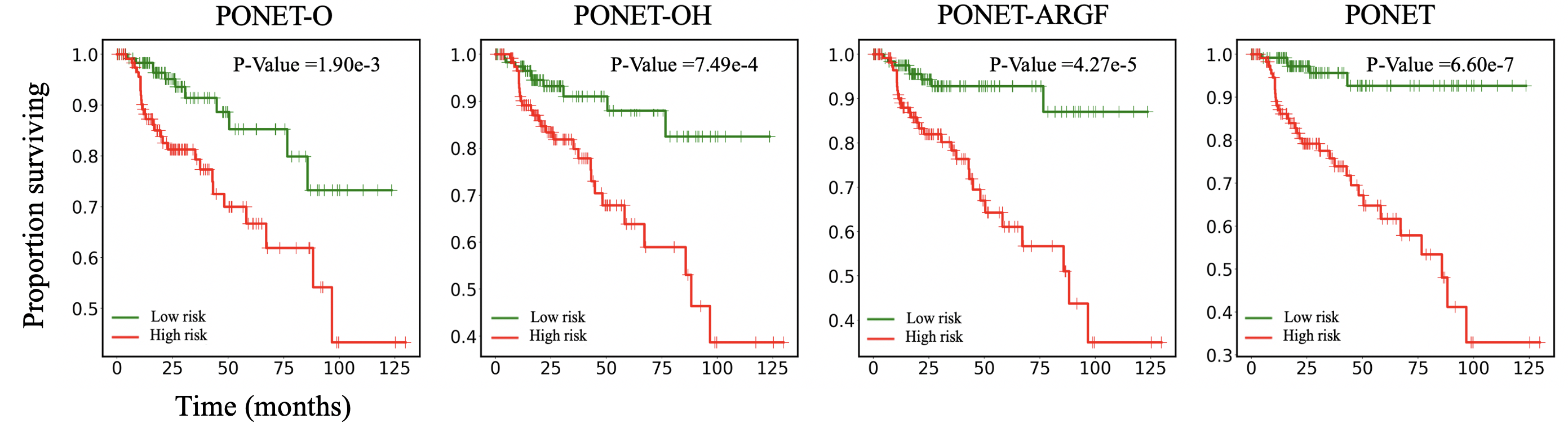}
    \caption{Kaplan-Meier analysis of patient stratification of low and high risk patients via four variations of PONET on TCGA-KIRP. Low and high risks are defined by the median 50$\%$ percentile of hazard predictions via each model prediction. Log-rank test was used to test for statistical significance in survival distributions between low and high risk patients.}
    \label{fig:km}
    \vskip -0.2in
\end{figure}

\textbf{Patient Stratification.} In visualizing the Kaplan-Meier survival curves of predicted
high risk and low risk patient populations, we plot four variations of PONET in Fig. \ref{fig:km}. PONET-ARGF represents the model that we use the hierarchical fusion strategy of ARGF in our pathway-informed PONET model. From the results, PONET enables easy separation of
patients into low and high risk groups with remarkably better stratification (P-Value = 6.60e-7) in comparison to the others.

\textbf{Complexity Comparison.} We compared PONET with Pathomic Fusion, GPDBN, and HFBSurv since both Pathomic Fusion and GPDBN are based on Kronecker product to fuse different modalities while GPDBN and HFBSurv modeled inter-modality and intra-modality relations which have similar consideration to our method. As illustrated in Table \ref{wrap-tab:1}, PONET has 2.8M (M = Million) trainable parameters, which is approximately 1.6\%, 3.4\%, and 900\% of the number of parameters of Pathomic Fusion, GPDBN, and HFBSurv. To assess the time complexity of PONET and the competitive methods, we calculate each method's floating-point operations per second (FLOPS) in testing. The results in Table \ref{wrap-tab:1} show that PONET needs 3.1G during testing, compared with 168G, 91G, and 0.5G in Pathomic Fusion, GPDBN, and HFBSurv. The main reason for fewer trainable parameters and the number of FLOPS lies in that PONET and HFBSurv perform multimodal fusion using the factorized bilinear model, and can significantly reduce the computational complexity and meanwhile obtain more favorable performance. PONET has one additional trimodal fusion which explains why it has more trainable parameters than HFBSurv.

\section{Conclusion}
In this study, we pioneer propose a novel biological pathway-informed hierarchical multimodal fusion model that integrates pathology image and genomic profile data for cancer prognosis. In comparison to previous works, PONET deeply mines the interaction from multimodal data by conducting unimodal, bimodal and trimodal fusion step by step. Empirically, PONET demonstrates the effectiveness of the model architecture and the pathway-informed network for superior predictive performance. 
Specifically, PONET provides insight on how to train biologically informed deep networks on multimodal biomedical data for biological discovery in clinic genomic contexts which will be useful for other problems in medicine that seek to combine heterogeneous data streams for understanding diseases and predicting response and resistance to treatment. 

\bibliography{iclr2023_conference}

\begin{thebibliography}{53}
\providecommand{\natexlab}[1]{#1}
\providecommand{\url}[1]{\texttt{#1}}
\expandafter\ifx\csname urlstyle\endcsname\relax
  \providecommand{\doi}[1]{doi: #1}\else
  \providecommand{\doi}{doi: \begingroup \urlstyle{rm}\Url}\fi

\bibitem[Al-Obaidy et~al.(2020)Al-Obaidy, Eble, Nassiri, Cheng, Eldomery,
  Williamson, Sakr, Gupta, Hassan, Idrees, et~al.]{ai-obaidy20}
Khaleel~I Al-Obaidy, John~N Eble, Mehdi Nassiri, Liang Cheng, Mohammad~K
  Eldomery, Sean~R Williamson, Wael~A Sakr, Nilesh Gupta, Oudai Hassan,
  Muhammad~T Idrees, et~al.
\newblock Recurrent kras mutations in papillary renal neoplasm with reverse
  polarity.
\newblock \emph{Modern Pathology}, 33(6):\penalty0 1157--1164, 2020.

\bibitem[Chan(2014)]{Chan14}
John~KC Chan.
\newblock The wonderful colors of the hematoxylin-eosin stain in diagnostic
  surgical pathology.
\newblock \emph{International Journal of Surgical Pathology}, 22(1):\penalty0
  12--32, 2014.

\bibitem[Cheerla \& Gevaert(2019)Cheerla and Gevaert]{cheerla19}
Anika Cheerla and Olivier Gevaert.
\newblock Deep learning with multimodal representation for pancancer prognosis
  prediction.
\newblock \emph{Bioinformatics}, 35:\penalty0 i446--i454, 2019.

\bibitem[Chen et~al.(2021)Chen, Lu, Weng, Chen, Williamson, Manz, Shady, and
  Mahmood]{chen21}
Richard~J Chen, Ming~Y Lu, Wei-Hung Weng, Tiffany~Y Chen, Drew~FK Williamson,
  Trevor Manz, Maha Shady, and Faisal Mahmood.
\newblock Multimodal co-attention transformer for survival prediction in
  gigapixel whole slide images.
\newblock \emph{IEEE/CVF International Conference on Computer Vision}, pp.\
  3995--4005, 2021.

\bibitem[Chen et~al.(2022{\natexlab{a}})Chen, Lu, Wang, Williamson, Rodig,
  Lindeman, and Mahmood]{chen22}
Richard~J Chen, Ming~Y Lu, Jingwen Wang, Drew~FK Williamson, Scott~J Rodig,
  Neal~I Lindeman, and Faisal Mahmood.
\newblock Pathomic fusion: An integrated framework for fusing histopathology
  and genomic features for cancer diagnosis and prognosis.
\newblock \emph{IEEE Transactions on Medical Imaging}, 41(4):\penalty0
  757--770, 2022{\natexlab{a}}.

\bibitem[Chen et~al.(2022{\natexlab{b}})Chen, Lu, Williamson, Chen, Lipkova,
  Noor, Shaban, Shady, Williams, Joo, et~al.]{chen23}
Richard~J Chen, Ming~Y Lu, Drew~FK Williamson, Tiffany~Y Chen, Jana Lipkova,
  Zahra Noor, Muhammad Shaban, Maha Shady, Mane Williams, Bumjin Joo, et~al.
\newblock Pan-cancer integrative histology-genomic analysis via interpretable
  multimodal deep learning.
\newblock \emph{arXiv:2108.02278}, 2022{\natexlab{b}}.

\bibitem[Cheng et~al.(2017)Cheng, Zhang, Han, Wang, Ye, Meng, Parwani, Han,
  Feng, and Huang]{cheng17}
Jun Cheng, Jie Zhang, Yatong Han, Xusheng Wang, Xiufen Ye, Yuebo Meng, Anil
  Parwani, Zhi Han, Qianjin Feng, and Kun Huang.
\newblock Integrative analysis of histopathological images and genomic data
  predicts clear cell renal cell carcinoma prognosis.
\newblock \emph{Cancer Research}, 77(21):\penalty0 e91--e100, 2017.

\bibitem[Cirillo et~al.(2017)Cirillo, Parnell, and Evelo]{cirillo17}
Elisa Cirillo, Laurence~D Parnell, and Chris~T Evelo.
\newblock A review of pathway-based analysis tools that visualize genetic
  variants.
\newblock \emph{Frontiers in Genetics}, 8(174), 2017.

\bibitem[Courtiol et~al.(2019)Courtiol, Maussion, Moarii, Pronier, Pilcer,
  Sefta, Manceron, Toldo, Zaslavskiy, Le~Stang, et~al.]{courtiol19}
Pierre Courtiol, Charles Maussion, Matahi Moarii, Elodie Pronier, Samuel
  Pilcer, Meriem Sefta, Pierre Manceron, Sylvain Toldo, Mikhail Zaslavskiy,
  Nolwenn Le~Stang, et~al.
\newblock Deep learning-based classification of mesothelioma improves
  prediction of patient outcome.
\newblock \emph{Nature Medicine}, 25(10):\penalty0 1519--1525, 2019.

\bibitem[Cox(1972)]{cox72}
David~R Cox.
\newblock Regression models and life-tables.
\newblock \emph{Journal of the Royal Statistical Society: Series B (Statistical
  Methodology)}, 34(2):\penalty0 187--202, 1972.

\bibitem[Elmarakeby et~al.(2021)Elmarakeby, Hwang, Arafeh, Crowdis, Gang, Liu,
  AlDubayan, Salari, Kregel, Richter, et~al.]{elmarakeby21}
Haitham~A Elmarakeby, Justin Hwang, Rand Arafeh, Jett Crowdis, Sydney Gang,
  David Liu, Saud~H AlDubayan, Keyan Salari, Steven Kregel, Camden Richter,
  et~al.
\newblock Biologically informed deep neural network for prostate cancer
  discovery.
\newblock \emph{Nature}, 598:\penalty0 348--352, 2021.

\bibitem[Fabregat et~al.(2020)Fabregat, Jupe, Matthews, Sidiropoulos,
  Gillespie, Garapati, Haw, Jassal, Korninger, May, et~al.]{Jassal20}
Antonio Fabregat, Steven Jupe, Lisa Matthews, Konstantinos Sidiropoulos, Marc
  Gillespie, Phani Garapati, Robin Haw, Bijay Jassal, Florian Korninger, Bruce
  May, et~al.
\newblock The reactome pathway knowledgebase.
\newblock \emph{Nucleic Acids Research}, 48(D1):\penalty0 D498--D503, 2020.

\bibitem[Grubb et~al.(2022)Grubb, Maganti, Krill-Burger, Fraser, Stransky,
  Radivoyevitch, Sarosiek, Vazquez, Jr., and Chakraborty]{grubb22}
Treg Grubb, Smruthi Maganti, John~Michael Krill-Burger, Cameron Fraser, Laura
  Stransky, Tomas Radivoyevitch, Kristopher~A. Sarosiek, Francisca Vazquez,
  William G.~Kaelin Jr., and Abhishek~A. Chakraborty.
\newblock A mesenchymal tumor cell state confers increased dependency on the
  bcl-xl anti-apoptotic protein in kidney cancer.
\newblock \emph{bioRxiv}, 2022.

\bibitem[Guo et~al.(2022)Guo, Jing, Li, and Liu]{guo22}
Jing-Yi Guo, Zuo-qian Jing, Xue-jie Li, and Li-yuan Liu.
\newblock Bioinformatic analysis identifying psmb 1/2/3/4/6/8/9/10 as
  prognostic indicators in clear cell renal cell carcinoma.
\newblock \emph{International Journal of Medical Sciences}, 19(5):\penalty0
  796--812, 2022.

\bibitem[Han et~al.(2018)Han, Kang, Joo, Lee, Oh, Choi, Ko, Je, Choi, and
  Song]{han18}
Zhezhu Han, Dongxu Kang, Yeonsoo Joo, Jihyun Lee, Geun-Hyeok Oh, Soojin Choi,
  Suwan Ko, Suyeon Je, Hye~Jin Choi, and Jae~J Song.
\newblock Tgf-beta downregulation-induced cancer cell death is finely regulated
  by the sapk signaling cascade.
\newblock \emph{Experimental and Molecular Medicine}, 50(12):\penalty0 162,
  2018.

\bibitem[Hao et~al.(2018)Hao, Kim, Kim, and Kang]{hao18}
Jie Hao, Youngsoon Kim, Tae-Kyung Kim, and Mingon Kang.
\newblock Pasnet: pathway-associated sparse deep neural network for prognosis
  prediction from high-throughput data.
\newblock \emph{BMC Bioinformatics}, 19:\penalty0 510, 2018.

\bibitem[Harrell et~al.(1982)Harrell, Califf, Pryor, Lee, and
  Rosati]{harrell82}
Frank~E Harrell, Robert~M Califf, David~B Pryor, Kerry~L Lee, and Robert~A
  Rosati.
\newblock Evaluating the yield of medical tests.
\newblock \emph{Journal of the American Medical Association}, 247(18):\penalty0
  2543--2546, 1982.

\bibitem[Hekler et~al.(2019)Hekler, Utikal, Enk, Solass, Schmitt, Klode,
  Schadendorf, Sondermann, Franklin, Bestvater, et~al.]{Hekler19}
Achim Hekler, Jochen~S Utikal, Alexander~H Enk, Wiebke Solass, Max Schmitt,
  Joachim Klode, Dirk Schadendorf, Wiebke Sondermann, Cindy Franklin, Felix
  Bestvater, et~al.
\newblock Deep learning outperformed 11 pathologists in the classification of
  histopathological melanoma images.
\newblock \emph{Europe Journal of Cancer}, 118:\penalty0 91--6, 2019.

\bibitem[Hou et~al.(2016)Hou, Samaras, Kurc, Gao, Davis, and Saltz]{hou16}
Le~Hou, Dimitris Samaras, Tahsin~M Kurc, Yi~Gao, James~E Davis, and Joel~H
  Saltz.
\newblock Patch-based convolutional neural network for whole slide tissue image
  classification.
\newblock In \emph{Proceedings of the IEEE/CVF Conference on Computer Vision
  and Pattern Recognition}, pp.\  2424--2433, 2016.

\bibitem[Hou et~al.(2020)Hou, Gupta, Van~Arnam, Zhang, Sivalenka, Samaras,
  Kurc, and Saltz]{Hou20}
Le~Hou, Rajarsi Gupta, John~S Van~Arnam, Yuwei Zhang, Kaustubh Sivalenka,
  Dimitris Samaras, Tahsin~M Kurc, and Joel~H Saltz.
\newblock Dataset of segmented nuclei in hematoxylin and eosin stained
  histopathology images of ten cancer types.
\newblock \emph{Scientific Data}, 7(1):\penalty0 185, 2020.

\bibitem[Iizuka et~al.(2020)Iizuka, Kanavati, Kato, Rambeau, Arihiro, and
  Tsuneki]{Iizuka20}
Osamu Iizuka, Fahdi Kanavati, Kei Kato, Michael Rambeau, Koji Arihiro, and
  Masayuki Tsuneki.
\newblock Deep learning models for histopathological classification of gastric
  and colonic epithelial tumours.
\newblock \emph{Scientific Report}, 10(1):\penalty0 1504, 2020.

\bibitem[Ilse et~al.(2018)Ilse, Tomczak, and Welling]{llse18}
Maximilian Ilse, Jakub Tomczak, and Max Welling.
\newblock Attention-based deep multiple instance learning.
\newblock In \emph{Proceedings of the 35th International Conference on Machine
  Learning}, volume~80, pp.\  2127--2136, 2018.

\bibitem[Jin et~al.(2014)Jin, Zuo, Su, Zhao, Yuan, Han, Zhao, Chen, and
  Rao]{jin14}
Lv~Jin, Xiao-Yu Zuo, Wei-Yang Su, Xiao-Lei Zhao, Man-Qiong Yuan, Li-Zhen Han,
  Xiang Zhao, Ye-Da Chen, and Shao-Qi Rao.
\newblock Pathway-based analysis tools for complex diseases: A review.
\newblock \emph{Genomics Proteomics Bioinformatics}, 12(5):\penalty0 210--220,
  2014.

\bibitem[Kampman et~al.(2018)Kampman, Barezi, Bertero, and Fung]{kampman18}
Onno Kampman, Elham~J Barezi, Dario Bertero, and Pascale Fung.
\newblock Investigating audio, video, and text fusion methods for end-to-end
  automatic personality prediction.
\newblock In \emph{Proceedings of the 56th Annual Meeting of the Association
  for Computational Linguistics (Volume 2: Short Papers)}, pp.\  606--611,
  2018.

\bibitem[Katzman et~al.(2018)Katzman, Shaham, Cloninger, Bates, Jiang, and
  Kluger]{katzman18}
Jared~L. Katzman, Uri Shaham, Alexander Cloninger, Jonathan Bates, Tingting
  Jiang, and Yuval Kluger.
\newblock Deepsurv: personalized treatment recommender system using a cox
  proportional hazards deep neural network.
\newblock \emph{BMC Medical Research Methology}, 18(24):\penalty0 187--202,
  2018.

\bibitem[Kim et~al.(2017)Kim, On, Lim, Kim, Ha, and Zhang]{kim17}
Jin-Hwa Kim, Kyoung-Woon On, Woosang Lim, Jeonghee Kim, Jung-Woo Ha, and
  Byoung-Tak Zhang.
\newblock Hadamard product for low-rank bilinear pooling.
\newblock In \emph{Proceedings of International Conference on Learning
  Representations}, pp.\  1--14, 2017.

\bibitem[Li et~al.(2022)Li, Wu, Li, and Wang]{li22}
Ruiqing Li, Xingqi Wu, Ao~Li, and Minghui Wang.
\newblock Hfbsurv: hierarchical multimodal fusion with factorized bilinear
  models for cancer survival prediction.
\newblock \emph{Bioinformatics}, 38(9):\penalty0 2587--2594, 2022.

\bibitem[Li et~al.(2015)Li, Nan, and Zhu]{li15}
Yanming Li, Bin Nan, and Ji~Zhu.
\newblock Multivariate sparse group lasso for the multivariate multiple linear
  regression with an arbitrary group structure.
\newblock \emph{Biometrics}, 71(2):\penalty0 354--63, 2015.

\bibitem[Lin et~al.(2017)Lin, Negulescu, Bulusu, Gibert, Delcros, Ducarouge,
  Rama, Gadot, Treilleux, Saintigny, et~al.]{lin17}
Shuheng Lin, Ana Negulescu, Sirisha Bulusu, Benjamin Gibert, Jean-Guy Delcros,
  Benjamin Ducarouge, Nicolas Rama, Nicolas Gadot, Isabelle Treilleux, Pierre
  Saintigny, et~al.
\newblock Non-canonical notch3 signalling limits tumour angiogenesis.
\newblock \emph{Nature Communications}, 8:\penalty0 16074, 2017.

\bibitem[Liu et~al.(2021)Liu, Shen, Lakshminarasimhan, Liang, Zadeh, and
  Morency]{liu18}
Zhun Liu, Ying Shen, Varun~Bharadhwaj Lakshminarasimhan, Paul~Pu Liang, Amir
  Zadeh, and Louis-Philippe Morency.
\newblock Efficient low-rank multimodal fusion with modality-specific factors.
\newblock In \emph{Proceedings of the 56th Annual Meeting of the Association
  for Computational Linguistics}, pp.\  2247--2256, 2021.

\bibitem[Lu et~al.(2021)Lu, Williamson, Chen, Chen, Barbieri, and
  Mahmood]{lu21}
Ming~Y Lu, Drew~FK Williamson, Tiffany~Y Chen, Richard~J Chen, Matteo Barbieri,
  and Faisal Mahmood.
\newblock Data-efficient and weakly supervised computational pathology on
  whole-slide images.
\newblock \emph{Nature Biomedical Engineering}, 5:\penalty0 555--570, 2021.

\bibitem[Mai et~al.(2020)Mai, Hu, and Xing]{mai20}
Sijie Mai, Haifeng Hu, and Songlong Xing.
\newblock Modality to modality translation: An adversarial representation
  learning and graph fusion network for multimodal fusion.
\newblock In \emph{Proceedings of the AAAI Conference on Artificial
  Intelligence}, pp.\  164--172, 2020.

\bibitem[Mallavarapu et~al.(2017)Mallavarapu, Kim, Oh, and Kang]{mallavarapu17}
Tejaswini Mallavarapu, Youngsoon Kim, Jung~Hun Oh, and Mingon Kang.
\newblock R-pathcluster: Identifying cancer subtype of glioblastoma multiforme
  using pathway-based restricted boltzmann machine.
\newblock In \emph{2017 IEEE International Conference on Bioinformatics and
  Biomedicine}, pp.\  1183--8, 2017.

\bibitem[Mobadersany et~al.(2018)Mobadersany, Yousefi, Amgad, Gutman,
  Barnholtz-Sloan, Vel{\'a}zquez~Vega, Brat, and Cooper]{mobadersany18}
Pooya Mobadersany, Safoora Yousefi, Mohamed Amgad, David~A Gutman, Jill~S
  Barnholtz-Sloan, Jos{\'e}~E Vel{\'a}zquez~Vega, Daniel~J Brat, and Lee~AD
  Cooper.
\newblock Predicting cancer outcomes from histology and genomics using
  convolutional networks.
\newblock \emph{Proceedings of the National Academy of Sciences},
  115(13):\penalty0 E2970--E2979, 2018.

\bibitem[Ning et~al.(2020)Ning, Pan, Chen, Xiao, Zhang, Luo, Wang, and
  Zhang]{ning20}
Zhenyuan Ning, Weihao Pan, Yuting Chen, Qing Xiao, Xinsen Zhang, Jiaxiu Luo,
  Jian Wang, and Yu~Zhang.
\newblock Integrative analysis of cross-modal features for the prognosis
  prediction of clear cell renal cell carcinoma.
\newblock \emph{Bioinformatics}, 36(9):\penalty0 2888--2895, 2020.

\bibitem[Nojavanasghari et~al.(2016)Nojavanasghari, Gopinath, Koushik,
  Baltru{\v{s}}aitis, and Morency]{Nojavanasghar16}
Behnaz Nojavanasghari, Deepak Gopinath, Jayanth Koushik, Tadas
  Baltru{\v{s}}aitis, and Louis-Philippe Morency.
\newblock Deep multimodal fusion for persuasiveness prediction.
\newblock In \emph{Proceedings of the 18th ACM International Conference on
  Multimodal Interaction}, pp.\  284--288, 2016.

\bibitem[Pan et~al.(2017)Pan, Li, Yang, Liu, Yang, Zhao, and Fan]{Pan17}
Xipeng Pan, Lingqiao Li, Huihua Yang, Zhenbing Liu, Jinxin Yang, Lingling Zhao,
  and Yongxian Fan.
\newblock Accurate segmentation of nuclei in pathological images via sparse
  reconstruction and deep convolutional networks.
\newblock \emph{Neurocomputing}, 15:\penalty0 88--99, 2017.

\bibitem[Poria et~al.(2016)Poria, Chaturvedi, Cambria, and Hussain]{poria16}
Soujanya Poria, Iti Chaturvedi, Erik Cambria, and Amir Hussain.
\newblock Convolutional mkl based multimodal emotion recognition and sentiment
  analysis.
\newblock In \emph{Proceedings of International Conference on Data Mining},
  pp.\  439--448, 2016.

\bibitem[Shan et~al.(2017)Shan, Tang, Qian, and Xia]{shan17}
Guang Shan, Tian Tang, Huijun Qian, and Yue Xia.
\newblock Expression of tiam1 and rac1 proteins in renal cell carcinoma and its
  clinical-pathological features.
\newblock \emph{International Journal of Clinical and Experimental Pathology},
  10(11):\penalty0 11114--11121, 2017.

\bibitem[Subramanian et~al.(2021)Subramanian, Syeda-Mahmood, and
  Do]{Subramania21}
Vaishnavi Subramanian, Tanveer Syeda-Mahmood, and Minh~N Do.
\newblock Multimodal fusion using sparse cca for breast cancer survival
  prediction.
\newblock In \emph{Proceedings of IEEE 18th International Symposium on
  Biomedical Imaging (ISBI)}, pp.\  1429--1432, 2021.

\bibitem[Sundararajan et~al.(2017)Sundararajan, Taly, and Yan]{Sundararajan17}
Mukund Sundararajan, Ankur Taly, and Qiqi Yan.
\newblock Axiomatic attribution for deep networks.
\newblock In \emph{Proceedings of the 34th International Conference on Machine
  Learning}, volume~70, pp.\  3319--3328, 2017.

\bibitem[Tenenbaum \& Freeman(2000)Tenenbaum and Freeman]{Tenenbaum00}
Joshua~B Tenenbaum and William~T Freeman.
\newblock Separating style and content with bilinear models.
\newblock \emph{Neural Computation}, 12(6):\penalty0 1247--1283, 2000.

\bibitem[Wang et~al.(2020)Wang, Guo, Zhao, Liu, Liu, Liu, and Guo]{wang19}
Chunyu Wang, Junling Guo, Ning Zhao, Yang Liu, Xiaoyan Liu, Guojun Liu, and
  Maozu Guo.
\newblock A cancer survival prediction method based on graph convolutional
  network.
\newblock \emph{IEEE Trans Nanobioscience}, 19(1):\penalty0 117--126, 2020.

\bibitem[Wang et~al.(2021{\natexlab{a}})Wang, Yang, Zhang, Wang, Zhang, Huang,
  Yang, and Han]{wang2021transpath}
Xiyue Wang, Sen Yang, Jun Zhang, Minghui Wang, Jing Zhang, Junzhou Huang, Wei
  Yang, and Xiao Han.
\newblock Transpath: Transformer-based self-supervised learning for
  histopathological image classification.
\newblock In \emph{International Conference on Medical Image Computing and
  Computer-Assisted Intervention}, pp.\  186--195. Springer,
  2021{\natexlab{a}}.

\bibitem[Wang et~al.(2021{\natexlab{b}})Wang, Li, Wang, and Li]{wang21}
Zhiqin Wang, Ruiqing Li, Minghui Wang, and Ao~Li.
\newblock Gpdbn: deep bilinear network integrating both genomic data and
  pathological images for breast cancer prognosis prediction.
\newblock \emph{Bioinformatics}, 27(18):\penalty0 2963--2970,
  2021{\natexlab{b}}.

\bibitem[W{\"o}llmer et~al.(2013)W{\"o}llmer, Weninger, Knaup, Schuller, Sun,
  Sagae, and Morency]{wollmer13}
Martin W{\"o}llmer, Felix Weninger, Tobias Knaup, Bj{\"o}rn Schuller, Congkai
  Sun, Kenji Sagae, and Louis-Philippe Morency.
\newblock Youtube movie reviews: Sentiment analysis in an audio-visual context.
\newblock \emph{IEEE Intelligent Systems}, 28(3):\penalty0 46--53, 2013.

\bibitem[Wulczyn et~al.(2020)Wulczyn, Steiner, Xu, Sadhwani, Wang,
  Flament-Auvigne, Mermel, Chen, Liu, and Stumpe]{wulczyn20}
Ellery Wulczyn, David~F Steiner, Zhaoyang Xu, Apaar Sadhwani, Hongwu Wang,
  Isabelle Flament-Auvigne, Craig~H Mermel, Po-Hsuan~Cameron Chen, Yun Liu, and
  Martin~C Stumpe.
\newblock Deep learning-based survival prediction for multiple cancer types
  using histopathology images.
\newblock \emph{Plos One}, 15(6):\penalty0 e0233678, 2020.

\bibitem[Yang et~al.(2017)Yang, Ji, Li, Fu, Jiang, and Meng]{yang17}
Chun-ming Yang, Shan Ji, Yan Li, Li-ye Fu, Tao Jiang, and Fan-dong Meng.
\newblock B-catenin promotes cell proliferation, migration, and invasion but
  induces apoptosis in renal cell carcinoma.
\newblock \emph{OncoTargets and Therapy}, 10:\penalty0 711--724, 2017.

\bibitem[Yu et~al.(2017)Yu, Yu, Fan, and Tao]{yu17}
Zhou Yu, Jun Yu, Jianping Fan, and Dacheng Tao.
\newblock Multi-modal factorized bilinear pooling with co-attention learning
  for visual question answering.
\newblock In \emph{IEEE International Conference on Computer Vision}, pp.\
  1839--1848, 2017.

\bibitem[Zadeh et~al.(2016)Zadeh, Zellers, Pincus, and Morency]{zadeh16}
Amir Zadeh, Rowan Zellers, Eli Pincus, and Louis-Philippe Morency.
\newblock Multimodal sentiment intensity analysis in videos: Facial gestures
  and verbal messages.
\newblock \emph{IEEE Intelligent Systems}, pp.\  82--88, 2016.

\bibitem[Zadeh et~al.(2017)Zadeh, Chen, Poria, Cambria, and Morency]{zadeh17}
Amir Zadeh, Minghai Chen, Soujanya Poria, Erik Cambria, and Louis-Philippe
  Morency.
\newblock Tensor fusion network for multimodal sentiment analysis.
\newblock In \emph{Proceedings of the 2017 Conference on Empirical Methods in
  Natural Language Processing}, pp.\  1103--1114, 2017.

\bibitem[Zadeh \& Schmid(2020)Zadeh and Schmid]{zadeh20}
Shekoufeh~Gorgi Zadeh and Matthias Schmid.
\newblock Bias in cross-entropy-based training of deep survival networks.
\newblock \emph{IEEE Transactions on Pattern Analysis and Machine
  Intelligence}, 43(9):\penalty0 3126--3137, 2020.

\bibitem[Zhang et~al.(2019)Zhang, Yin, Pan, Cao, Han, Gao, Gao, Pan, and
  Feng]{zhang19}
Qiang Zhang, Xiujuan Yin, Zhiwei Pan, Yingying Cao, Shaojie Han, Guojun Gao,
  Zhiqin Gao, Zhifang Pan, and Weiguo Feng.
\newblock Identification of potential diagnostic and prognostic biomarkers for
  prostate cancer.
\newblock \emph{Oncology Letters}, 18(4):\penalty0 4237--4245, 2019.

\end{thebibliography}
\bibliographystyle{iclr2023_conference}
\clearpage
\appendix
\section{Data}
Table 3 in Appendix \ref{appendix:data} shows the number of patients with matched different data modalities: WSI (Whole slide image), CNV (Copy number), MUT (Mutation), RNA (RNA-Seq gene expression). For each TCGA dataset and each patient we have preprocessed data dimensions  $\mathbf{d}_{\mathrm{g}} \in \mathbb{R}^{1 \times 2000}$
(RNA), $\mathbf{d}_{\mathrm{c}} \in \mathbb{R}^{1 \times 227}$ (CNV + MUT), and $\mathbf{d}_{\mathrm{p}} \in \mathbb{R}^{1 \times 32}$ (WSI) which will be used for our multimodal fusion.
\begin{table} \label{appendix:data}
\caption {{TCGA Data Feature Alignment Summary}}
\centering
\begin{tabular}{l|rrrr|rrr}
\toprule
{} &   WSI &   CNV &   MUT &   RNA &  WSI+CNV+MUT &  WSI+MUT+RNA &   ALL \\
Cancer Type  &       &       &       &       &              &              &       \\
\midrule
BLCA     &   454 &   443 &   452 &   450 &          441 &          448 &   \textbf{437} \\
KIRC     &   517 &   509 &   357 &   514 &          352 &          355 &   \textbf{350} \\
KIRP     &   294 &   291 &   286 &   293 &          284 &          285 &   \textbf{284} \\
LUAD     &   528 &   522 &   523 &   522 &          519 &          519 &   \textbf{515} \\
LUSC     &   505 &   502 &   489 &   503 &          486 &          487 &   \textbf{484} \\
PAAD     &   208 &   201 &   187 &   195 &          187 &          180 &   \textbf{180} \\
\bottomrule
\end{tabular}
\end{table}
\section{Methods}

\subsection{C-Index}\label{appendix:cindx}
We use concordance-index (C-index) \citep{harrell82} to measure the performance of survival models. It evaluates the model by measuring the concordance of the ranking of predicted harzards with the true survival time of patients. The range of the $\mathrm{C}$-index is $[0,1]$, and larger values indicate better performance with a random guess leading to a $\mathrm{C}$-index of $0.5$. 

\subsection{WSI representation learning} \label{appendix:wsi}
It has been shown that the WSI visual representations extracted by self-supervised learning methods on histopathological images are more accurate and transferable than the supervised baseline models on domain-irrelevant datasets such as ImageNet. In this work, a pre-trained Vision Transformer (ViT) model \citep{wang2021transpath} that is trained on a large histopathological image dataset has been utilized for tile feature extraction. The model is composed of two main neural networks that learn from each other, i.e., student and teacher networks. Parameters of the teacher model $\theta_{t}$ are updated using the student network with parameter $\theta_{s}$ using the update rule represented in Eq. (\ref{eq:online_target}).


\begin{equation}\label{eq:online_target}
\theta_{t} \leftarrow \tau \theta_{t}+(1-\tau) \theta_{s}
\end{equation}

Two different views of a given input H\&E image ${x}$, uniformly selected from the training set $\mathcal{I}$, are generated using random augmentations, i.e., ${u}$, ${v}$. Then, student and teacher models generate two different visual representations according to ${u}$ and ${v}$ as $y_{1}=f^{\theta_{s}}\left({u}\right)$ and $\hat{y}_{2}=f^{\theta_{t}}\left({v}\right)$, respectively. Finally, the generated visual representations are transformed into latent space using linear projection as $p_{1}=g^{\theta_{s}}\left(g^{\theta_{s}}\left(y_{1}\right)\right)$ and $\hat{z_{2}}=g^{\theta_{t}}\left(\hat{y}_{2}\right)$ for student and teacher networks, respectively. Similarly, feeding ${v}$ and ${u}$ to student and teacher networks leads to $y_{2}=f^{\theta_{s}}\left({v}\right), \hat{y_{1}}=f^{\theta_{t}}\left({u}\right), p_{2}=g^{\theta_{s}}\left(g^{\theta_{s}}\left(y_{2}\right)\right)$ and $\hat{z_{1}}=g^{\theta_{t}}\left(\hat{y_{1}}\right)$. Finally, the symmetric objective function $L_{l o s s}$ is optimized through minimizing the $\ell_{2}-\text{norm}$ distance between student and teacher as Eq. (\ref{eq:sym_loss})

\begin{equation}\label{eq:sym_loss}
L_{l o s s}=\frac{1}{2} L\left(p_{1}, \hat{z}_{2}\right)+\frac{1}{2} L\left(p_{2}, \hat{z}_{1}\right)
\end{equation}

where $L(p, z)=-\frac{p}{\|p\|_{2}} \cdot \frac{z}{\|z\|_{2}}$ and $\| \cdot \|_{2}$ represents $\ell_{2}-\text {norm}$.

\subsection{Sparse network feature interpretation} \label{appendix:Intgradient}
We use the Integrated Gradients attribution algorithm to rank the features in all layers. Inspired by PNET \citep{elmarakeby21},  to reduce the bias introduced by over-annotation of certain nodes (nodes that are members of too many pathways), we adjusted the Integrated Gradients scores using a graph informed function $f$ that considers the connectivity of each node. The importance score of each node $i$, $C_i^l$ is divided by the node degree $d_i^l$ if the node degree is larger than the mean of node degrees plus $5 \sigma$ where $\sigma$ is the standard deviation of node degrees.

$$
d_i^l=fan-in_i^l+fan-out_i^l
$$

adjusted  $C_i^l=f(x)= \begin{cases}\frac{C_i^l}{d_i^l}, & d_i^l>\mu+5 \sigma \\ C_i^l, & \text { otherwise }\end{cases}$

\subsection{Co-attention based pathway visualization} \label{appendix:pathway_visu}
After we got the ranking of top genes and pathways, we adopted the co-attention survival model (MCAT) \citep{chen21} to show the spatial visualization of genomic features. We trained MACT on all our TCGA datasets, and MACT learns how WSI patches attend to genes when predicting patient survival. We define each WSI patch representation and pathway genomic features as $H_{bag}$ and $G_{bag}$. The genomic features are the gene list values from the top pathways of each TCGA dataset. The model uses $G_{bag} \in \mathbb{R}^{N\times d_g}$ to guide the feature aggregation of $H_{bag} \in \mathbb{R}^{N\times d_p}$ into a clustered set of gene-guided visual concepts $\widehat{H}_{\text {bag }} \in \mathbb{R}^{N \times d_p}$ , $d_g$ and $d_p$ represents the dimension for the pathway (number of genes involved in the pathway) and patch.  Through the following mapping:
$$\operatorname{CoAttn}_{G \rightarrow H}(G, H)=\operatorname{softmax}\left(\frac{Q K^{\top}}{\sqrt{d_p}}\right)$$
$$=\operatorname{softmax}\left(\frac{\mathbf{W}_q G H^{\top} \mathbf{W}_s^{\top}}{\sqrt{d_p}}\right) \mathbf{W}_v H \rightarrow A_{\text {coattn }} \mathbf{W}_v H \rightarrow \widehat{H}$$

where $\mathbf{W}_q, \mathbf{W}_s, \mathbf{W}_v \in \mathbb{R}^{d_p \times d_p}$ are trainable weight matrices multiplied to the queries $G_{\text {bag }}$ and key-value pair ($H_{\text {bag }}, H_{\text {bag }}$), and $A_{\text {coattn }} \in \mathbb{R}^{N \times M}$ is the co-attention matrix for computing the weighted average of $H_{\text {bag }}$. Here, $M$ represents the number of patches in one slide, and $N$ represents the number of pathways (We trained the top four pathways, so $N =4$ in our study).

\section{Experiments}

\subsection{Network architecture}\label{section:network} 

\textbf{Sparse network for gene}: The final gene expression embedding is  $\mathbf{h}_{\mathrm{g}} \in \mathbb{R}^{1 \times 50}$.

\textbf{Pathology network}: The slide level image feature representation is passed through an image embedding layer and encodes the embedding as $\mathbf{h}_{\mathrm{p}} \in \mathbb{R}^{ 1 \times 50}$.

\textbf{CNV + MUT network}: Similarly as the pathology network, the patient level CNV + MUT feature representation is passed through an FC embedding layer and encodes the embedding as $\mathbf{h}_{\mathrm{c}} \in \mathbb{R}^{1 \times 50}$.

\subsection{Experimental Details} \label{appendix:training_details}
\begin{figure}[t!] \label{appendix:cindex_tune}
    \centering
    \includegraphics[width=0.7\textwidth]{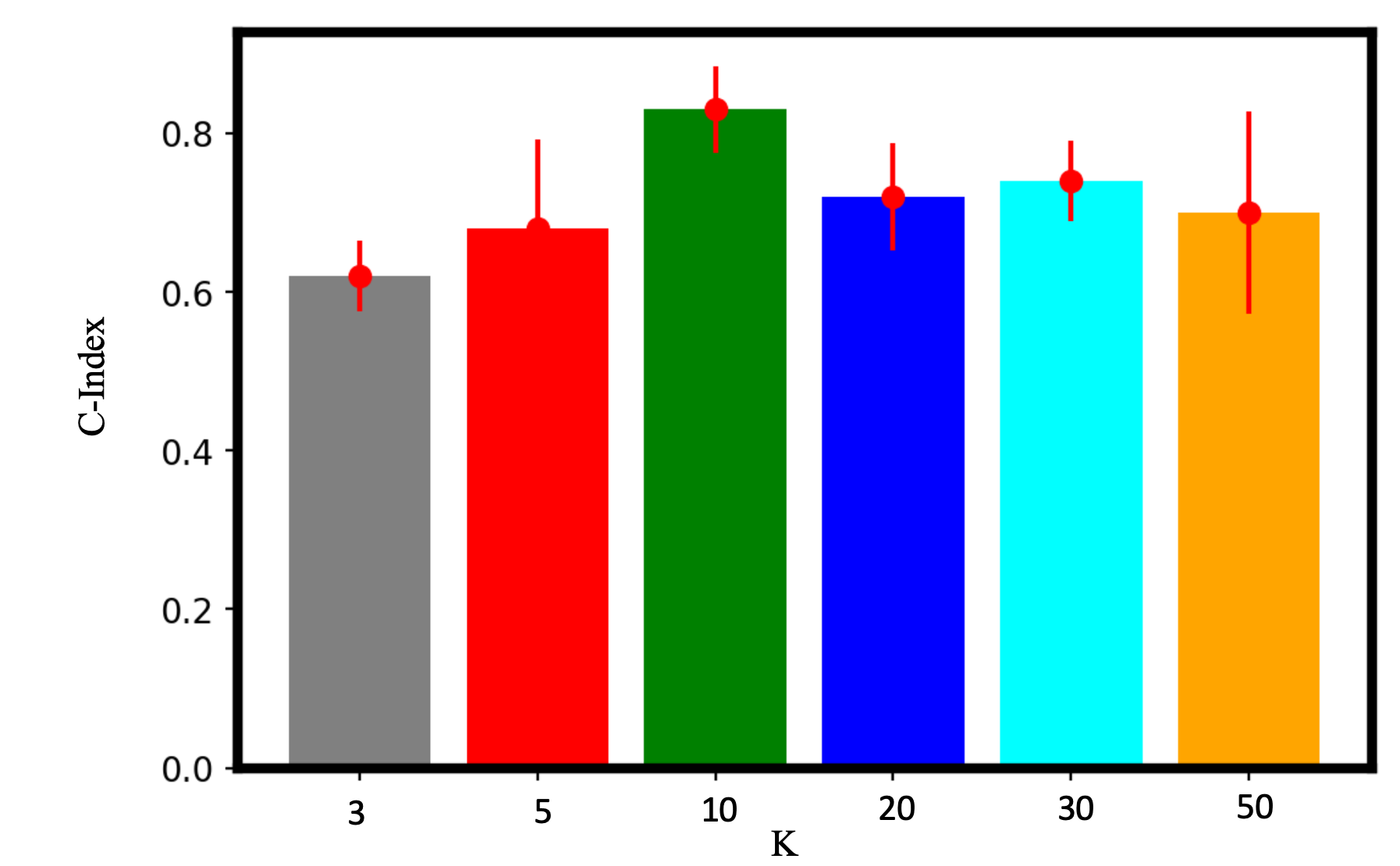}
    \caption{C-Index value under K = 3, 5, 10, 20, 30, 50 for TCGA-KIRP. The mean value and standard deviation for 5-fold cross-validation are plotted.}
    \label{fig:cindex_tune}
\end{figure}

\begin{figure}[t!] \label{appendix:train_test_loss}
    \centering
    \includegraphics[width=1\textwidth]{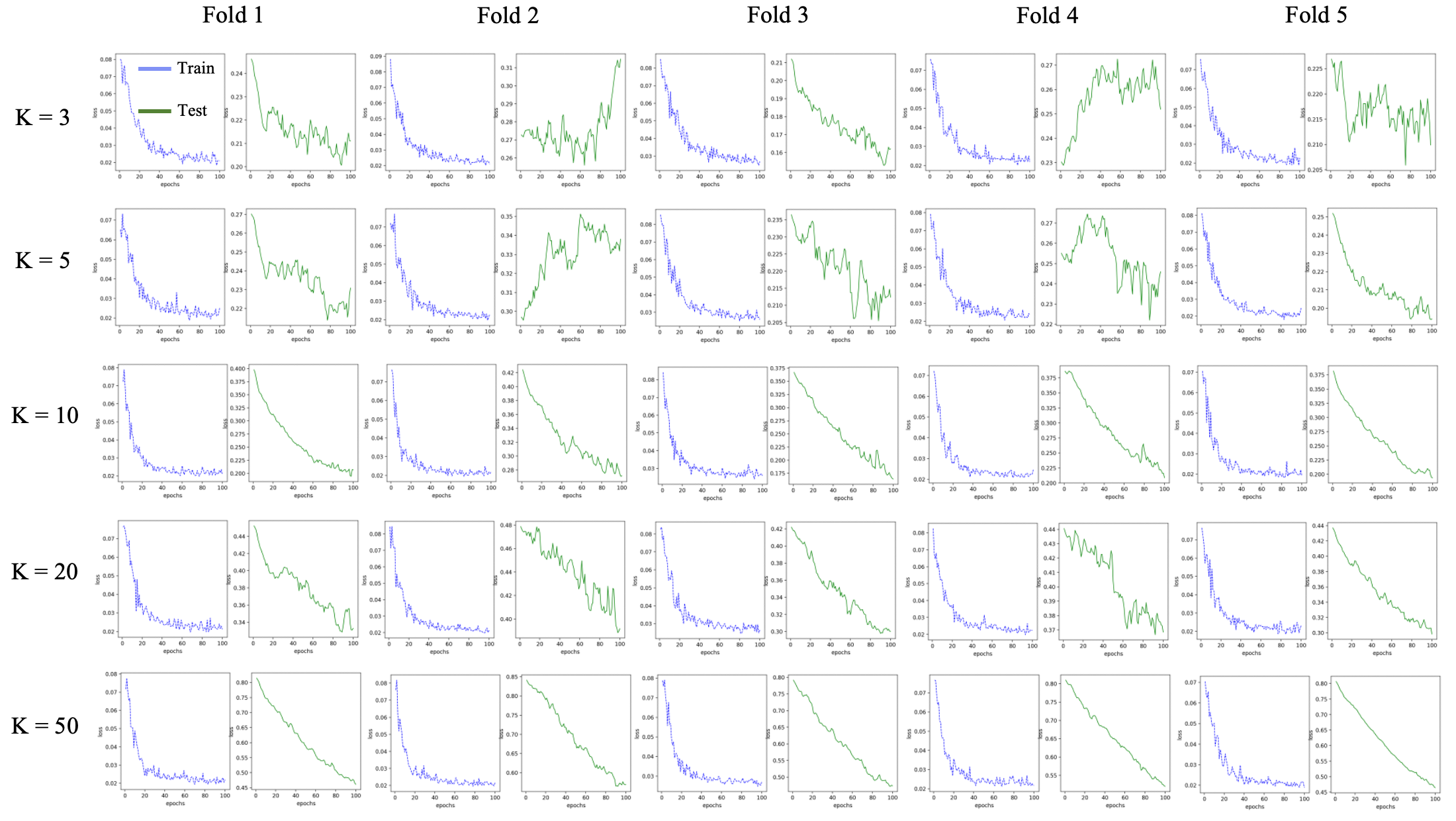}
    \caption{Train and test loss for TCGA-KIRP under K = 3, 5, 10, 20, 50 for  5-fold cross-validation.}
    \label{fig:train_test_loss}
\end{figure}

\textbf{PONET}. The latent dimensionality of the factorized matrices $k$ is a very important tuning parameter. We tune $k=[3, 5, 10, 20, 30, 50]$ based on the testing C-index value (Appendix Fig. \ref{fig:cindex_tune}) and the loss of training and testing plot (Appendix Fig. \ref{fig:train_test_loss}) for each dataset. We choose $k$ to maximize the C-index value and also it should have stable convergence in both training and testing loss. For example,  we choose $k=10$ in TCGA-KIRP for the optimized results. We can see that in Appendix Fig. \ref{fig:cindex_tune} the testing loss is quite volatile when $k$ is less than 10. Similarly, we choose $k=[20, 10, 20, 20, 10]$ for TCGA-BLCA, TCGA-KIRC, TCGA-LUAD, TCGA-LUSC, and TCGA-PAAD, respectively.  

The learning rate and the regularization hyperparameter $\lambda$ for the Cox partial likelihood loss are also tunable parameters. The model is trained with Adam optimizer. For each training/testing pair, we first empirically preset the learning rate to 1.2e-4 as a starting point for a grid search during training, the optimal learning rate is determined through the 5-fold cross-validation on the training set, C-index was used for the performance metric. After that, the model is trained on all the training sets and evaluated on the testing set. We use 2e-3 through the experiments for $\lambda$. The batch size is set to 16, and the epoch is 100. During the training process, we carefully observe the training and testing loss for convergence (Figure 4 in Appendix \ref{appendix:train_test_loss}). The server used for experiments is NVIDIA GeForce RTX 2080Ti GPU. \\
\textbf{CoxPH}. We only include the age and gender for the survival prediction. Using CoxPHFitter from lifelines \footnote{https://github.com/CamDavidsonPilon/lifelines}. \\
\textbf{DeepSurv} \footnote{https://github.com/czifan/DeepSurv.pytorch}. We concatenate preprocessed pathological image features, gene expression, and copy number + mutant data in a vector to train the DeepSurv model. L2 reg = 10.0, dropout = 0.4, hidden layers sizes = [25, 25], learning rate = 1e-05, learning rate decay = 0.001, momentum = 0.9. \\
\textbf{Pathomic Fusion} \footnote{https://github.com/mahmoodlab/PathomicFusion}. We use the pathomicSurv model which takes our preprocessed image feature, gene expression, and copy number + mutation as model input.  $k$ = 20, Learning rate is 2e-3, weight decay is 4e-4. The batch size is 16, and the epoch is 100. Drop out rate is 0.25. \\
\textbf{GPDBN} \footnote{https://github.com/isfj/GPDBN}. The learning rate is 2e-3, the batch size is 16, the weight decay is 1e-6, the dropout rate is 0.3, and the epoch is 100. \\

\textbf{HFBSurv} \footnote{https://github.com/Liruiqing-ustc/HFBSurv}. The learning rate is set to 1e-3, the batch size is 16, $\lambda$  = 3e-3, weight decay is 1e-6, and the epoch is 100.

\clearpage
\subsection{Additional Results} \label{appendix:additional_results}


	%
	
 


\begin{figure}[hbt!] 
    \centering
    \includegraphics[width=1\textwidth]{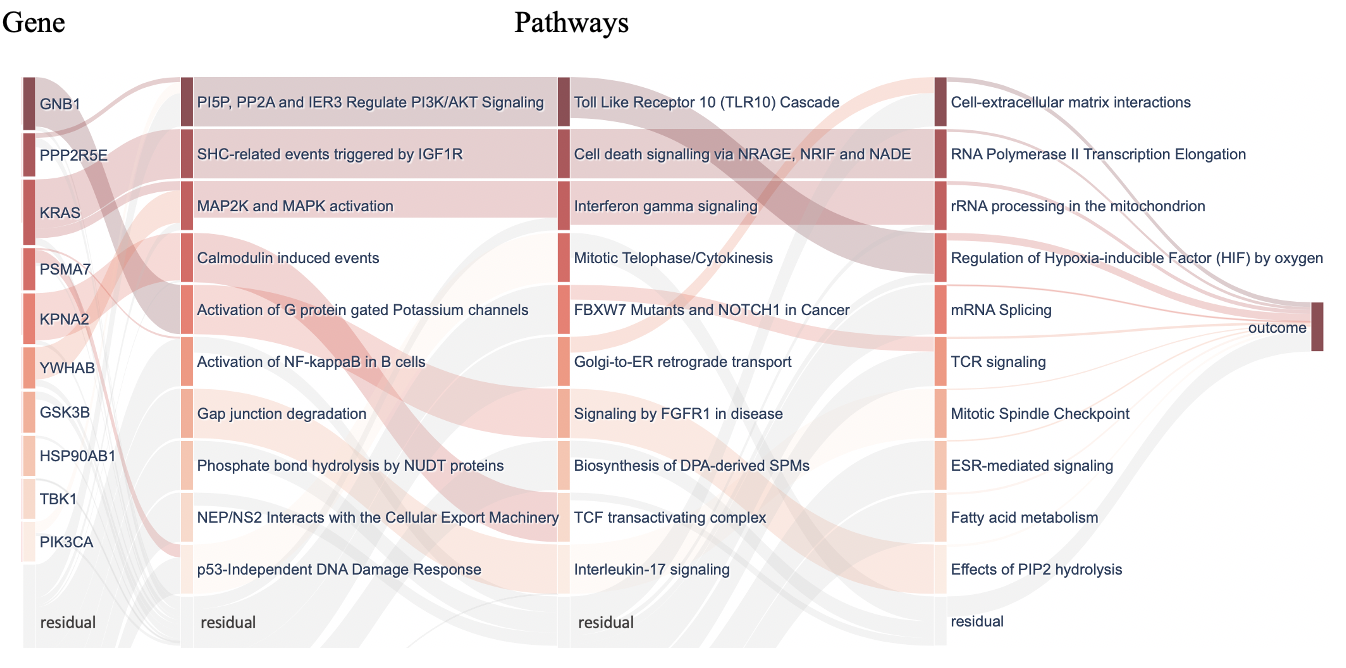}
    \caption{Inspecting and interpreting PONET on TCGA-BLCA. Sankey diagram visualization of the inner layers of PONET shows the estimated relative importance of different nodes in each layer. Nodes in the first layer
represent genes; the next layers represent pathways; and
the final layer represents the model outcome. Different layers are linked by weights.  Nodes with darker colors are
more important, while transparent nodes represent the residual importance of
undisplayed nodes in each layer.}
    \label{fig:blca_sankey}
\end{figure}

\begin{figure}[hbt!]
    \centering
    \includegraphics[width=1\textwidth]{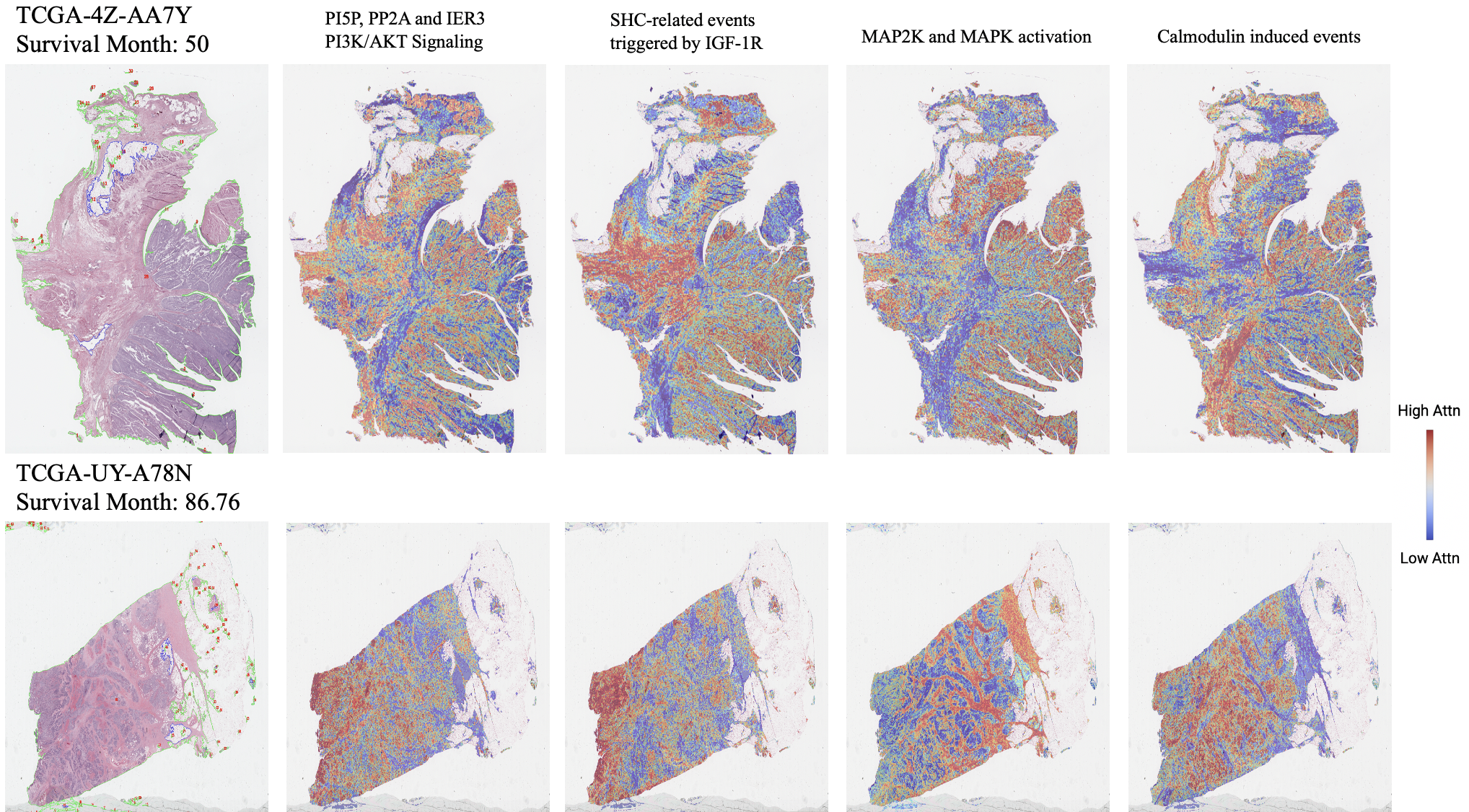}
    \caption{Co-attention visualization of top 4 ranked pathways in two cases of TCGA-BLCA.}
    \label{fig:pathway_img_blca}
\end{figure}

\clearpage

\begin{figure}[hbt!] 
    \centering
    \includegraphics[width=1\textwidth]{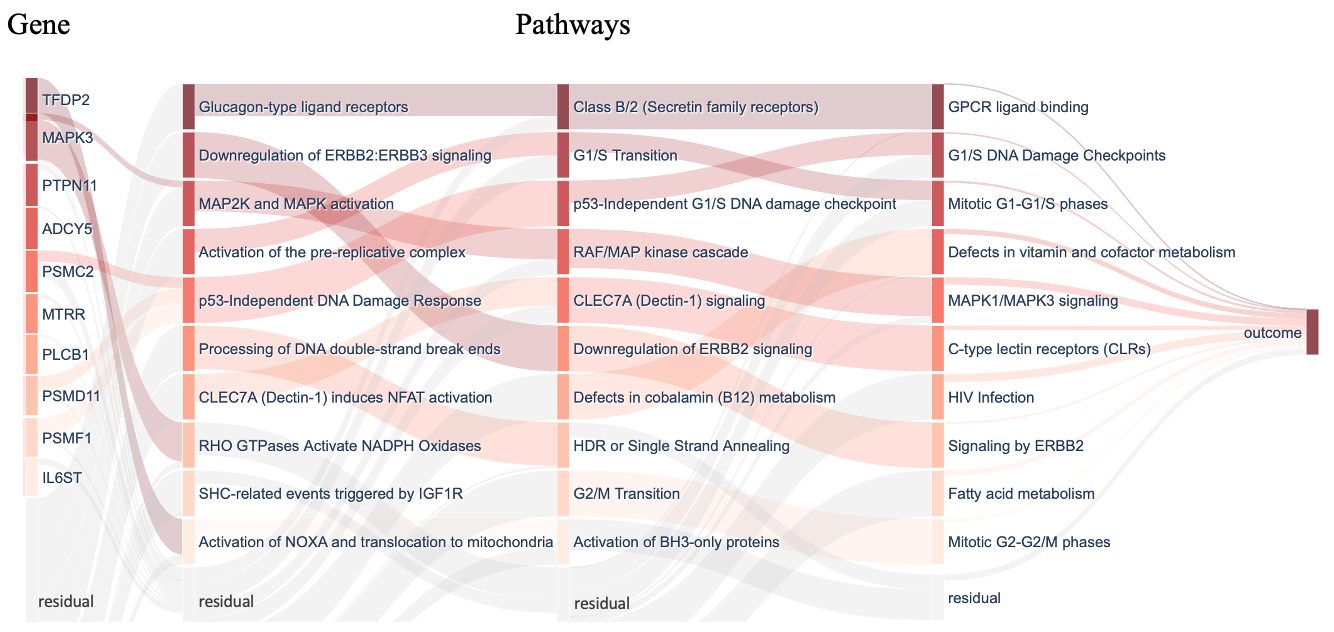}
    \caption{Inspecting and interpreting PONET on TCGA-KIRC. Sankey diagram visualization of the inner layers of PONET shows the estimated relative importance of different nodes in each layer. Nodes in the first layer
represent genes; the next layers represent pathways; and
the final layer represents the model outcome. Different layers are linked by weights. Nodes with darker colors are
more important, while transparent nodes represent the residual importance of
undisplayed nodes in each layer.}
    \label{fig:kirc_sankey}
\end{figure}

\begin{figure}[hbt!]
    \centering
    \includegraphics[width=1\textwidth]{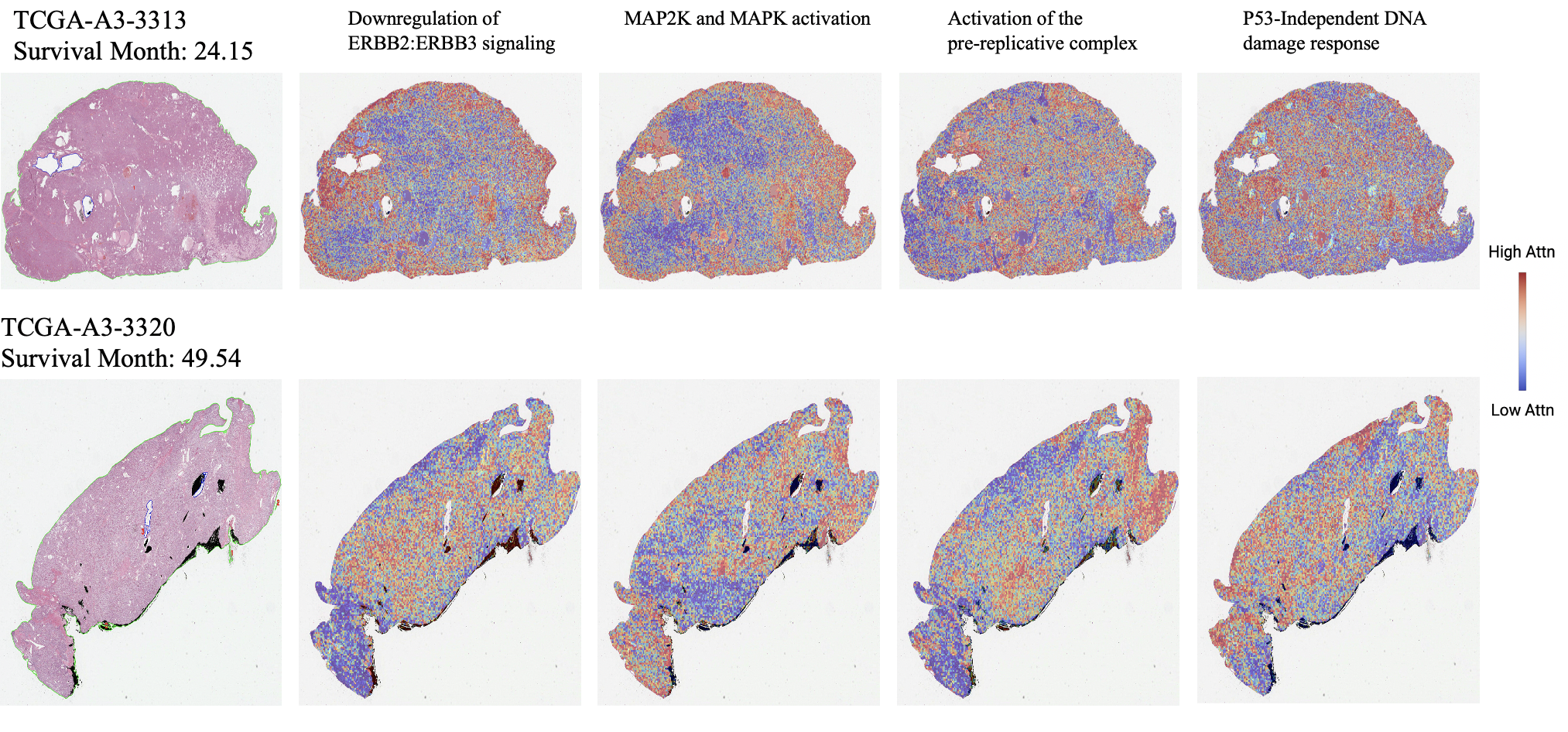}
    \caption{Co-attention visualization of top 4 ranked pathways in two cases of TCGA-KIRC.}
    \label{fig:pathway_img_kirc}
\end{figure}

\clearpage

\begin{figure}[hbt!] 
    \centering
    \includegraphics[width=1\textwidth]{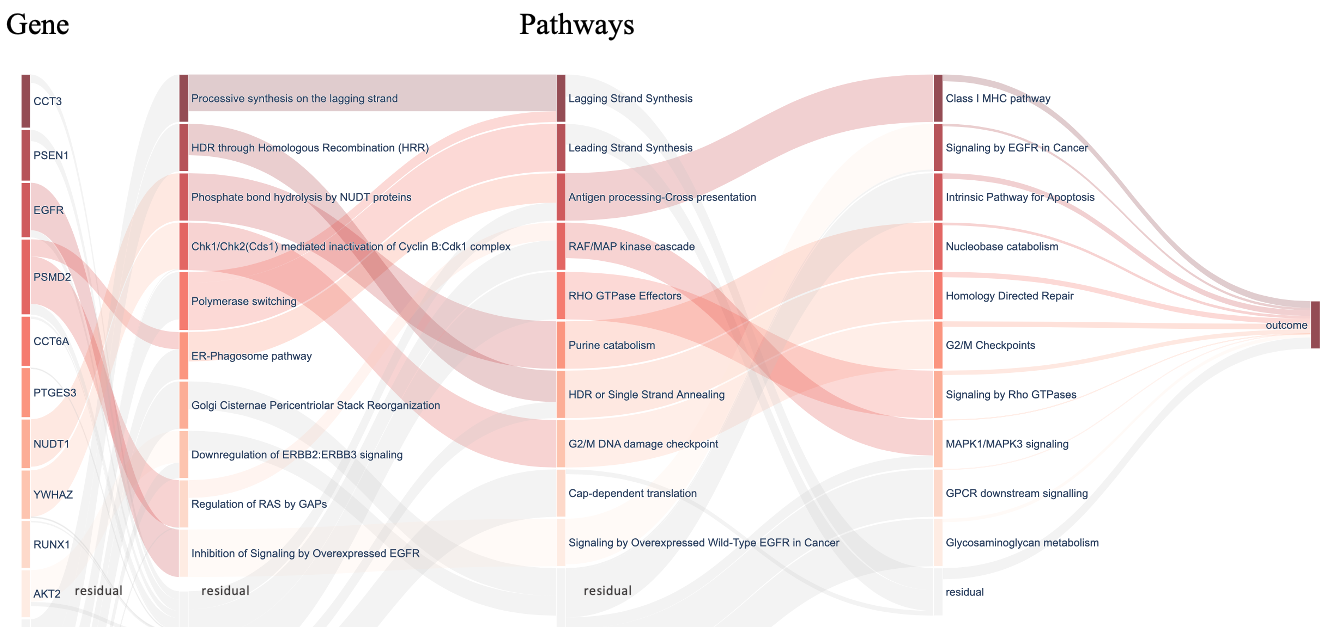}
    \caption{Inspecting and interpreting PONET on TCGA-LUAD. Sankey diagram visualization of the inner layers of PONET shows the estimated relative importance of different nodes in each layer. Nodes in the first layer
represent genes; the next layers represent pathways; and
the final layer represents the model outcome. Different layers are linked by weights. Nodes with darker colors are
more important, while transparent nodes represent the residual importance of
undisplayed nodes in each layer.}
    \label{fig:luad_sankey}
\end{figure}

\begin{figure}[hbt!]
    \centering
    \includegraphics[width=1\textwidth]{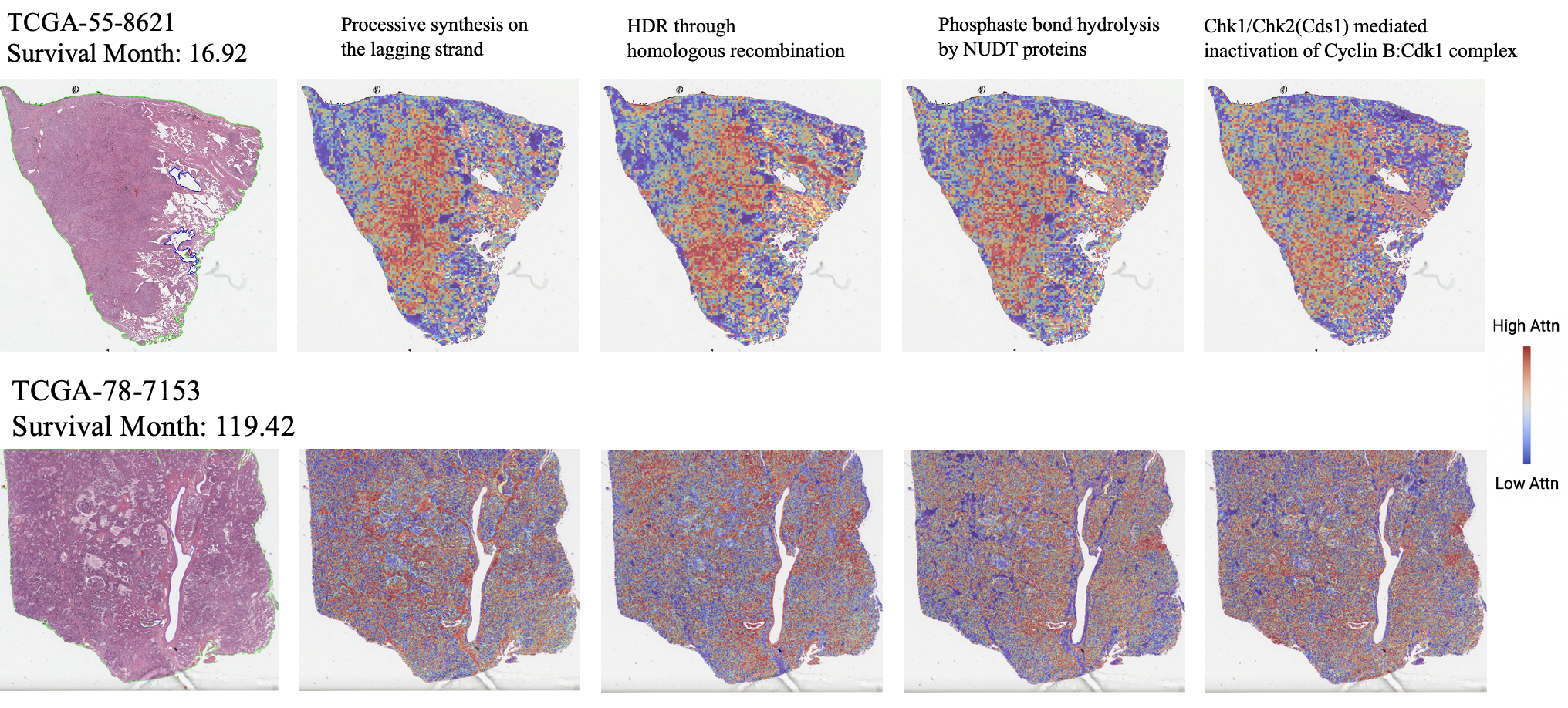}
    \caption{Co-attention visualization of top 4 ranked pathways in two cases of TCGA-LUAD.}
    \label{fig:pathway_img_luad}
\end{figure}

\clearpage

\begin{figure}[hbt!] 
    \centering
    \includegraphics[width=1\textwidth]{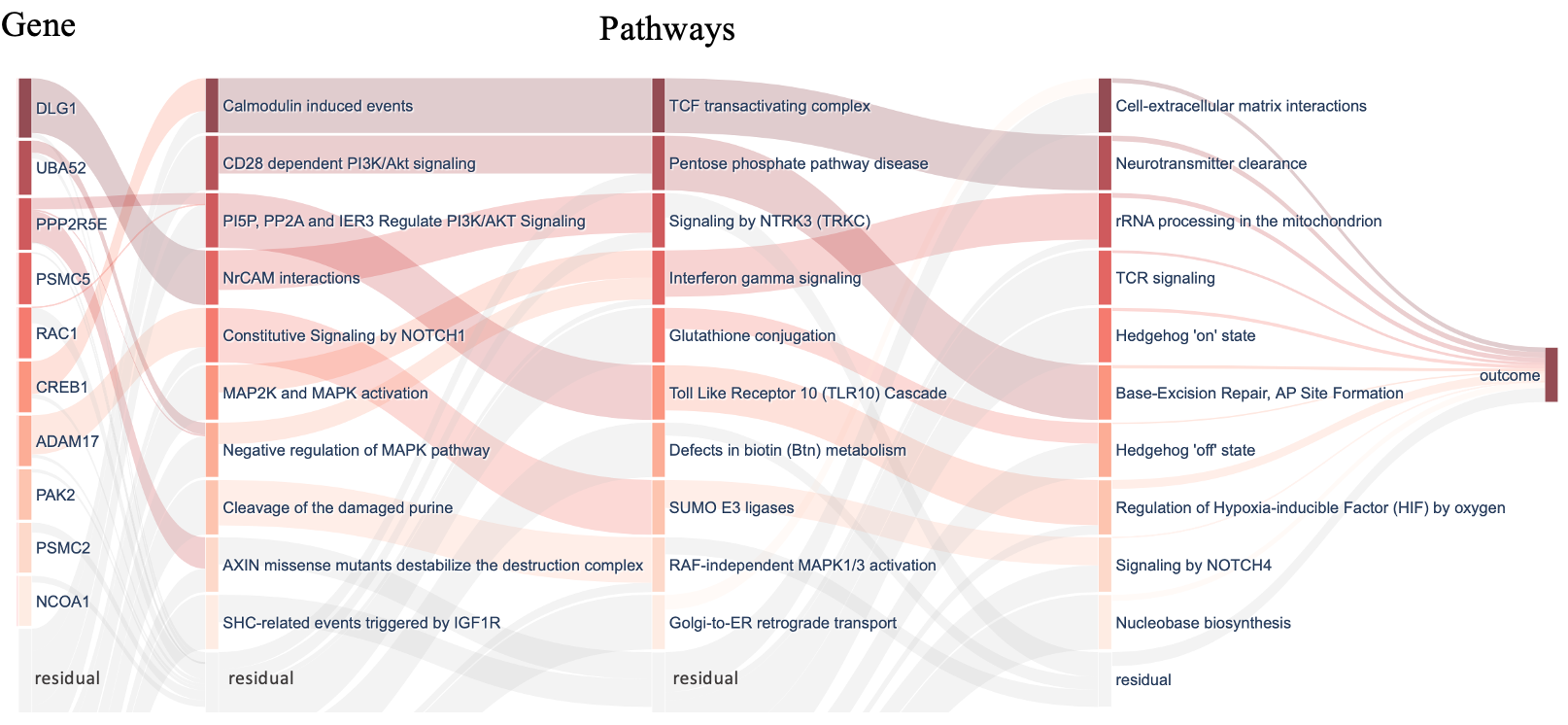}
    \caption{Inspecting and interpreting PONET on TCGA-LUSC. Sankey diagram visualization of the inner layers of PONET shows the estimated relative importance of different nodes in each layer. Nodes in the first layer
represent genes; the next layers represent pathways; and
the final layer represents the model outcome. Different layers are linked by weights.  Nodes with darker colors are
more important, while transparent nodes represent the residual importance of
undisplayed nodes in each layer.}
    \label{fig:lusc_sankey}
\end{figure}

\begin{figure}[hbt!]
    \centering
    \includegraphics[width=1\textwidth]{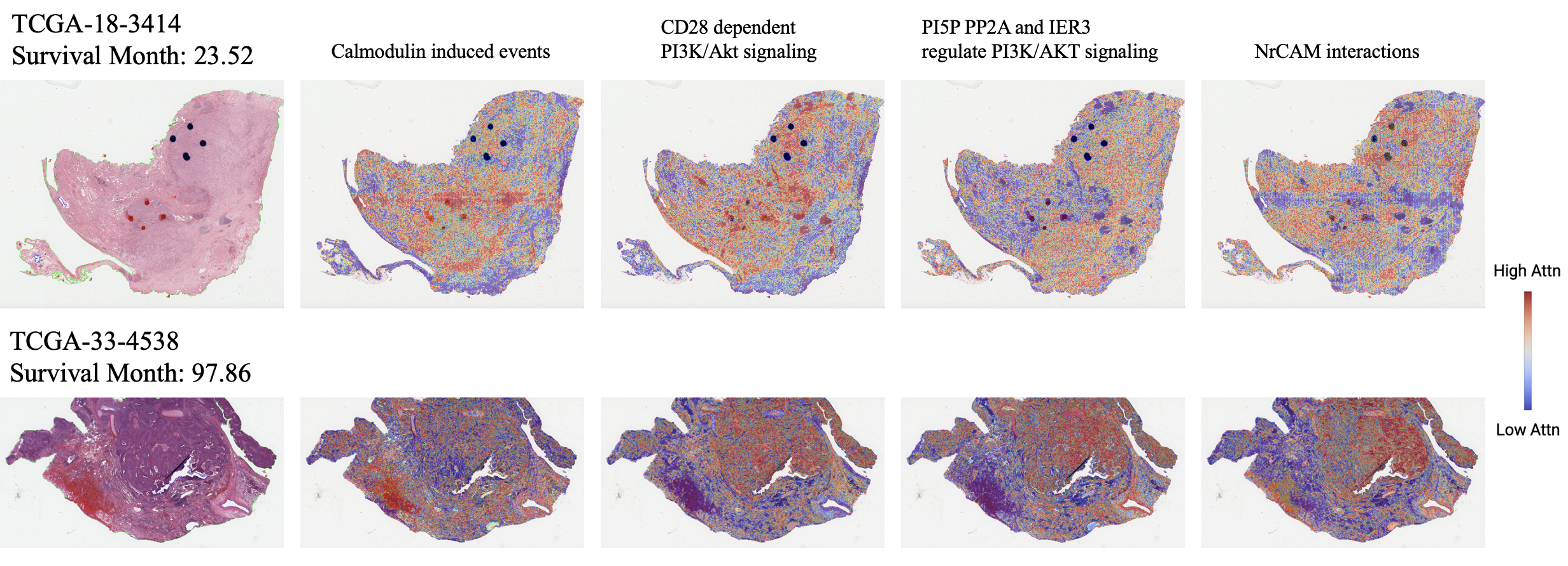}
    \caption{Co-attention visualization of top 4 ranked pathways in two cases of TCGA-LUSC.}
    \label{fig:pathway_img_lusc}
\end{figure}

\clearpage
\begin{figure}[hbt!] 
    \centering
    \includegraphics[width=1\textwidth]{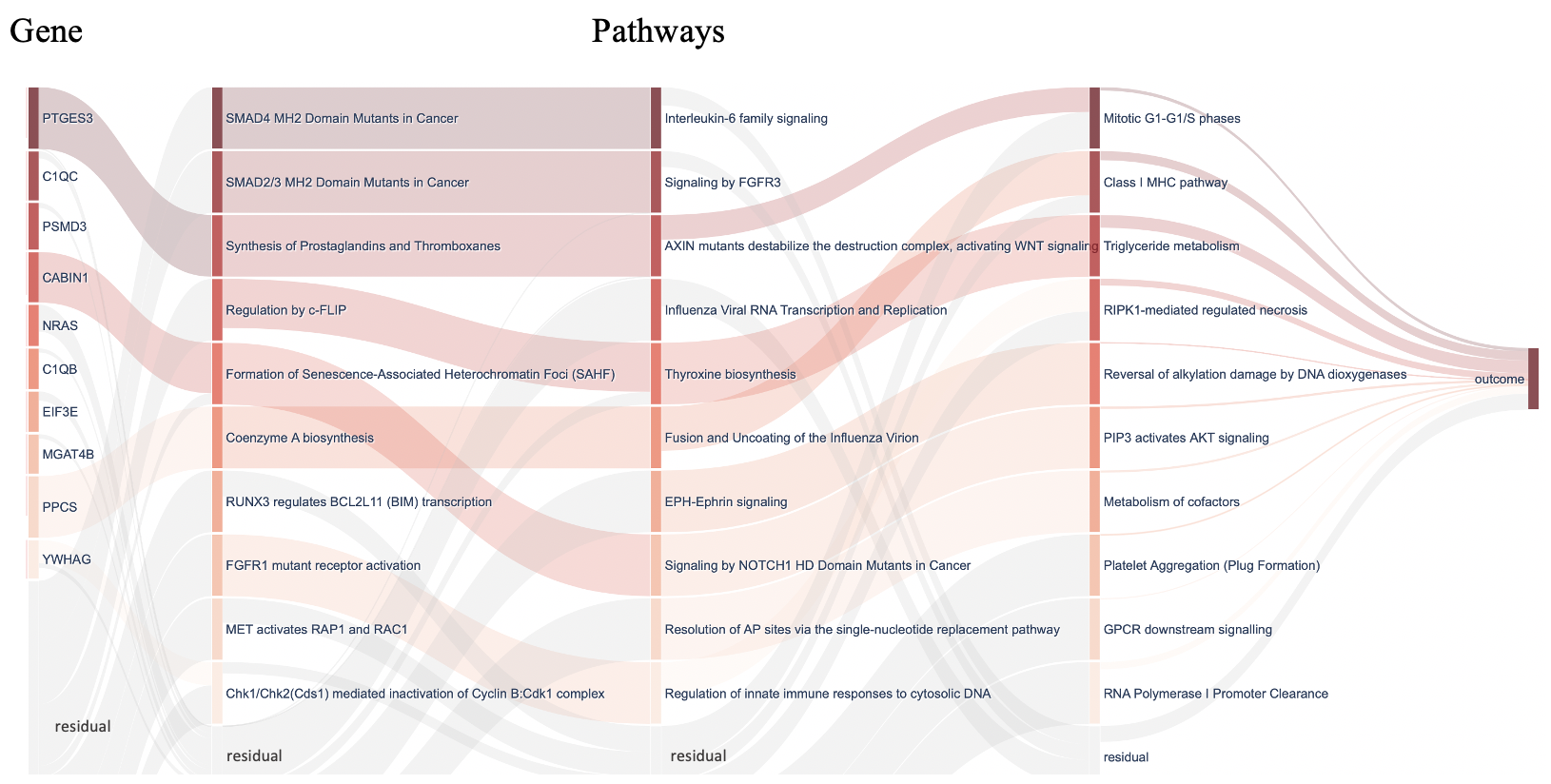}
    \caption{Inspecting and interpreting PONET on TCGA-PAAD. Sankey diagram visualization of the inner layers of PONET shows the estimated relative importance of different nodes in each layer. Nodes in the first layer
represent genes; the next layers represent pathways; and
the final layer represents the model outcome. Different layers are linked by weights. Nodes with darker colors are
more important, while transparent nodes represent the residual importance of
undisplayed nodes in each layer.}
    \label{fig:paad_sankey}
\end{figure}

\begin{figure}[hbt!]
    \centering
    \includegraphics[width=1\textwidth]{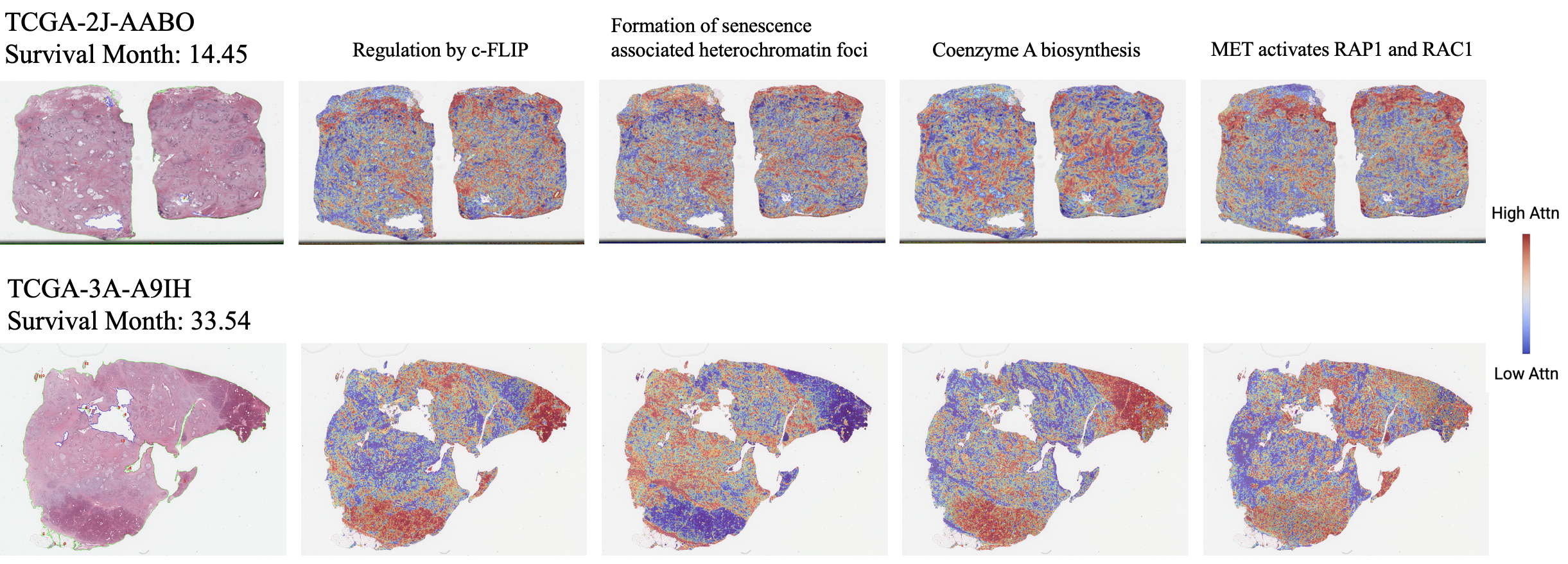}
    \caption{Co-attention visualization of top 4 ranked pathways in two cases of TCGA-PAAD.}
    \label{fig:pathway_img_paad}
\end{figure}

\end{document}